\documentclass[ twocolumn,superscriptaddress, amsmath, showpacs,
tightenlines,twocolumn,twoside,pra]{revtex4}%
\usepackage{amsfonts}
\usepackage{amstext}
\usepackage{amsmath}
\usepackage{amssymb}
\usepackage{latexsym}
\usepackage{graphicx,epstopdf}
\usepackage{ulem}
\usepackage[colorlinks,linkcolor=blue,anchorcolor=blue,citecolor=blue]%
{hyperref}
\usepackage{times}
\usepackage{subfigure}
\usepackage{color}
\usepackage{dcolumn}
\usepackage{graphicx}
\usepackage{soul}
\usepackage{xcolor}
\usepackage{framed}%
\providecommand{\U}[1]{\protect\rule{.1in}{.1in}}
\begin{document}
\title{Vortex-Meissner phase transition induced by two-tone-drive-engineered
artificial gauge potential in the fermionic ladder constructed by
superconducting qubit circuits}
\author{Yan-Jun Zhao }
\affiliation{Faculty of Information Technology, College of Microelectronics, Beijing
University of Technology, Beijing, 100124, People's Republic of China}
\author{Xun-Wei Xu }
\affiliation{Department of Applied Physics, East China Jiaotong University, Nanchang
330013, China}
\author{Hui Wang}
\affiliation{Center for Emergent Matter Science, RIKEN, 2-1 Hirosawa, Wako-shi, Saitama
351-0198, Japan}
\author{Yu-xi Liu}
\affiliation{Institute of Microelectronics, Tsinghua University, Beijing 100084, China}
\affiliation{Frontier Science Center for Quantum Information, Beijing, China}
\author{Wu-Ming Liu}
\affiliation{Beijing National Laboratory for Condensed Matter Physics, Institute of
Physics, Chinese Academy of Sciences, Beijing 100190, China}
\keywords{superconducting qubit circuit, quantum simulation, artifical gauge potential,
vortex phase, Meissner phase, chiral current}
\pacs{}

\begin{abstract}
We propose to periodically modulate the onsite energy via two-tone drives,
which can be furthermore used to engineer artificial gauge potential. As an
example, we show that the fermionic ladder model penetrated with effective
magnetic flux can be constructed by superconducting flux qubits using such
two-tone-drive-engineered artificial gauge potential. In this superconducting
system, the single-particle ground state can range from vortex phase to
Meissner phase due to the competition between the interleg coupling strength
and the effective magnetic flux. We also present the method to experimentally
measure the chiral currents by the single-particle Rabi oscillations between
adjacent qubits. In contrast to previous methods of generating artifical gauge
potential, our proposal does not need the aid of auxiliary couplers and in
principle remains valid only if the qubit circuit maintains enough
anharmonicity. The fermionic ladder model with effective magnetic flux can
also be interpreted as one-dimensional spin-orbit-coupled model, which thus
lay a foundation towards the realization of quantum spin Hall effect.

\end{abstract}
\revised{\today}

\startpage{1}
\endpage{ }
\maketitle

%\author{Wu-Ming Liu}
%\affiliation{Beijing National Laboratory for Condensed Matter Physics, Institute of
%Physics, Chinese Academy of Sciences, Beijing 100190, China}
%\affiliation{School of Physical Sciences, University of Chinese Academy of Sciences,
%Beijing 100190, China}

\section{Introduction}%

%TCIMACRO{\TeXButton{schematic diagram}{\begingroup\begin{figure*}[ptb]
%\includegraphics[width=0.6\textwidth, clip]{fig1.eps}\caption{(color
%online). Ladder model constructed by X-shape flux qubits with the gradiometer structure
%which can cancel out some common flux noise penetrated through the two symmetric loops.
%The Josephson junctions, flux qubit loop, readout resonator, classical flux bias
%are colored in gray, red, green, and blue, respectively.
%The Josephson energy for the big and small Josephson junctions are respectively $E_J$
%and $\alpha E_J$, where $\frac{1}{2}<\alpha
%<1$ should be satisfied to guarantee the nonlineartiy of the flux qubit and meanwhile,
%suppress the intercell tunnelling.
%Meanwhile, the flux qubits are coupled to
%their nearest neighbours with X-shape mutual inductances that are mostly determined by the nearest edge on the loop.
%The microwave coplanar waveguide resonator (CPW) is shorted at the terminal near the flux qubit loop
%such that only inductive coupling is present. The flux qubit loop is designed like
%a cross such that different coupling terms can be well separately to minimize the crosstalk.}
%\label{fig:schematic diagram}
%\end{figure*}
%\endgroup}}%
%BeginExpansion
\begingroup\begin{figure*}[ptb]
\includegraphics[width=0.6\textwidth, clip]{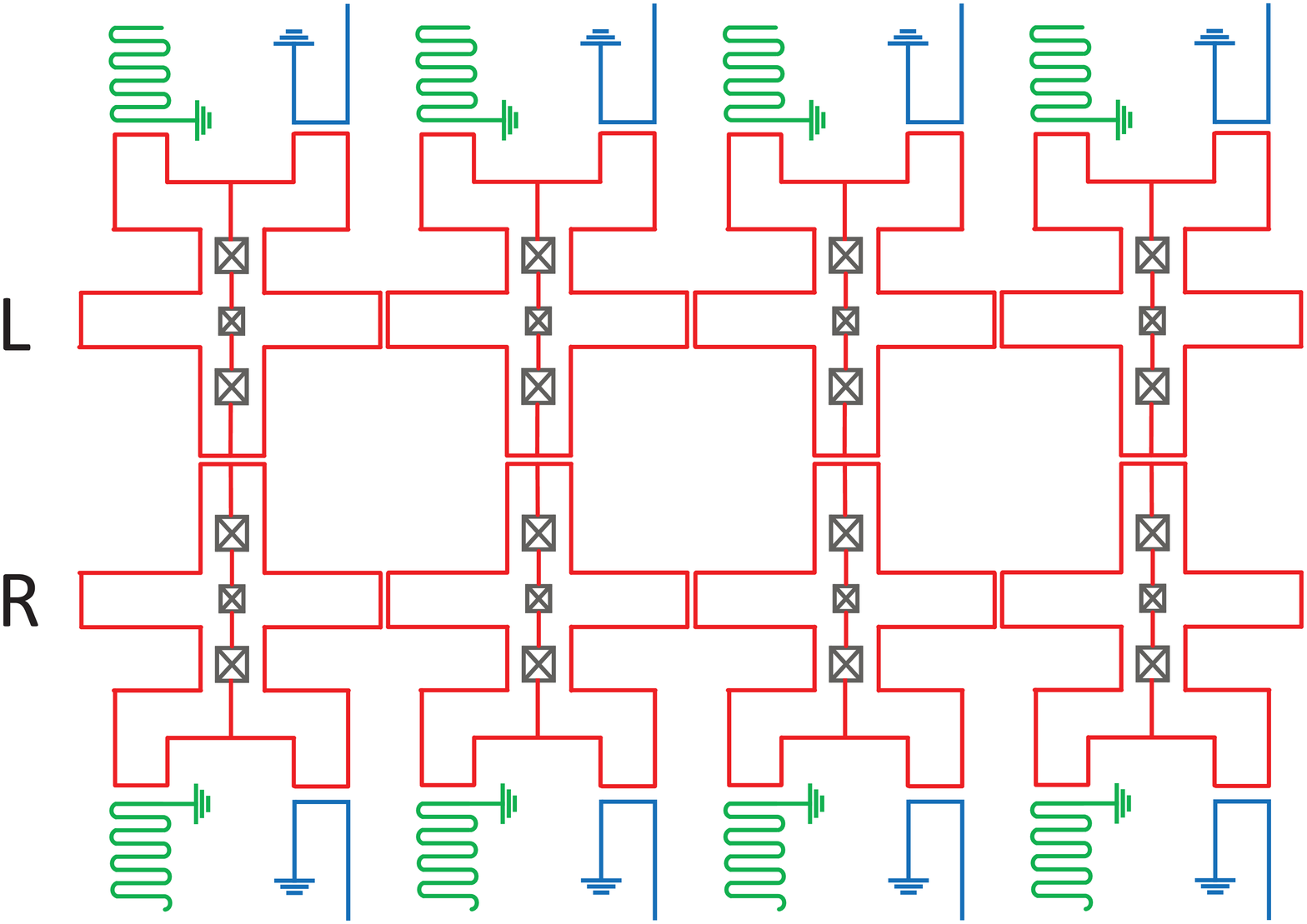}\caption{(color
online). Ladder model constructed by X-shape flux qubits with the gradiometer structure
which can cancel out some common flux noise penetrated through the two symmetric loops.
The Josephson junctions, flux qubit loop, readout resonator, classical flux bias
are colored in gray, red, green, and blue, respectively.
The Josephson energy for the big and small Josephson junctions are respectively $E_J$
and $\alpha E_J$, where $\frac{1}{2}<\alpha
<1$ should be satisfied to guarantee the nonlineartiy of the flux qubit and meanwhile,
suppress the intercell tunnelling.
Meanwhile, the flux qubits are coupled to
their nearest neighbours with X-shape mutual inductances that are mostly determined by the nearest edge on the loop.
The microwave coplanar waveguide resonator (CPW) is shorted at the terminal near the flux qubit loop
such that only inductive coupling is present. The flux qubit loop is designed like
a cross such that different coupling terms can be well separately to minimize the crosstalk.}
\label{fig:schematic diagram}
\end{figure*}
\endgroup
%EndExpansion

Gauge potential is a core ingredient of the electromagnetic interaction in
electrodynamics%
%TCIMACRO{\TeXButton{TeX field}{~}}%
%BeginExpansion
~%
%EndExpansion
\cite{Jackson1999Book}, standard model in particle physics%
%TCIMACRO{\TeXButton{TeX field}{~}}%
%BeginExpansion
~%
%EndExpansion
\cite{Gaillard1999RMP}, and even the topological phenomena in condensed matter
physics%
%TCIMACRO{\TeXButton{TeX field}{~}}%
%BeginExpansion
~%
%EndExpansion
\cite{Hasan2010RMP}. However, the behaviours of microscopic particles in gauge
potentials are rather difficult to study in natural systems, due to their
well-known low controllability. Representatively, for example, strong magnetic
field is experimentally challenging to generate for electrons in solid
systems. Therefore, engineering effective gauge potential in artificial
quantum platform stands a wise option in order to access higher tunability.
Superconducting qubit circuits%
%TCIMACRO{\TeXButton{TeX field}{~}}%
%BeginExpansion
~%
%EndExpansion
\cite{Makhlin2001RMP,You2005Phys.Today,Wendin2007LTP,Clarke2008Nature,Schoelkopf2008Nature,Buluta2011RPP,You2001Nature,Xiang2013RMP,Gu2017PR}%
, which inherit the advantages of microwave circuits in flexibility of design,
convenience of scaling up, and maturation of controlling technology, have
recently won great celebrity in simulating the motions of microscopic
particles placed in gauge potentials. In superconducting qubit circuits,
photons play the role of carriers, which, in contrast to electrons, will cause
no backaction onto the artificial gauge potential due to the charge neutrality.

The engineering of artificial gauge potential (mainly the effective magnetic
flux) in superconducting qubit circuits greatly depends on the nonlinearity of
Josephson junctions in auxiliary couplers%
%TCIMACRO{\TeXButton{TeX field}{~}}%
%BeginExpansion
~%
%EndExpansion
\cite{Koch2010PRA,Nunnenkamp2011NJP,Marcos2013PRL,Yang2016PRA,Roushan2017NP}.
In this manner, chiral Fock-state transfer%
%TCIMACRO{\TeXButton{TeX field}{~}}%
%BeginExpansion
~%
%EndExpansion
\cite{Koch2010PRA}, multiparticle spectrum modulated by effective magnetic
flux in Jaynes-Cummings model%
%TCIMACRO{\TeXButton{TeX field}{~}}%
%BeginExpansion
~%
%EndExpansion
\cite{Nunnenkamp2011NJP}, condensed-matter and high-energy physics phenomena
in quantum-link model%
%TCIMACRO{\TeXButton{TeX field}{~}}%
%BeginExpansion
~%
%EndExpansion
\cite{Marcos2013PRL}, and flat band in the Lieb lattice%
%TCIMACRO{\TeXButton{TeX field}{~}}%
%BeginExpansion
~%
%EndExpansion
\cite{Yang2016PRA} have been theoretically studied. In experiment,
effective-magnetic-flux-induced chiral currents of single photon and
single-photon vacancy have been respectively observed in one-photon and
two-photon states%
%TCIMACRO{\TeXButton{TeX field}{~}}%
%BeginExpansion
~%
%EndExpansion
\cite{Roushan2017NP}. By contrast, in cold atom systems, artificial gauge
potentials are usually engineered using periodically-modulated onsite energy%
%TCIMACRO{\TeXButton{TeX field}{~}}%
%BeginExpansion
~%
%EndExpansion
\cite{Aidelsburger2011PRL,Aidelsburger2013PRL,Miyake2013PRL,Atala2014NP}. This
has motivated the similar proposal of engineering artificial gauge potentials
via periodically modulating the Josephson energy of the transmon qubit circuit%
%TCIMACRO{\TeXButton{TeX field}{~}}%
%BeginExpansion
~%
%EndExpansion
\cite{Alaeian2019PRA}, which however maintains valid only in small
anharmonicity regime. To remedy this drawback, we propose to modulate the
onsite energy of the coupled qubit chain with two-tone drives. This method can
in principle be applied to a superconducting qubit circuit with any nonzero
anharmonicity, which can thus simulate fermions rather than bosons as in Ref.%
%TCIMACRO{\TeXButton{TeX field}{~}}%
%BeginExpansion
~%
%EndExpansion
\cite{Alaeian2019PRA}. Besides, nonlinearity is known to be a key factor for
demonstrating quantum phenomena%
%TCIMACRO{\TeXButton{TeX field}{~}}%
%BeginExpansion
~%
%EndExpansion
\cite{Wendin2007LTP}. Thus, periodically modulating the energy of the qubit
circuit with better anharmonicity is significant for exploring nonequilibrium
quantum physics.

Meanwhile, thanks to the recent experimental progress in the integration scale
of superconducting qubit circuits%
%TCIMACRO{\TeXButton{TeX field}{~}}%
%BeginExpansion
~%
%EndExpansion
\cite{Barends2014Nature,Zheng2017PRL,Song2017PRL,Gong2019PRL}, the\ quantum
simulation research based on superconducting qubit circuits is now advancing
from single or several qubits%
%TCIMACRO{\TeXButton{TeX field}{~}}%
%BeginExpansion
~%
%EndExpansion
\cite{Leak2007Science,Berger2012PRB,Berger2013PRA,Schroer2014PRL,Zhang2017PRA,Roushan2014Nature,Flurin2017PRX,Ramasesh2017PRL,Tan2017npj,Tan2018PRL,Zhong2016PRL,Guo2019PRapp}
towards multiple qubits%
%TCIMACRO{\TeXButton{TeX field}{~}}%
%BeginExpansion
~%
%EndExpansion
\cite{Roushan2017NP,Nunnenkamp2011NJP,Mei2015PRA,Yang2016PRA,Tangpanitanon2016PRL,Gu2017arXiv,Roushan2017Science,Xu2018PRL,Yan2019Science,Ye2019PRL,Arute2019Nature}%
. However, most experiments are yet confined to the chain structure (one
dimension) currently%
%TCIMACRO{\TeXButton{TeX field}{~}}%
%BeginExpansion
~%
%EndExpansion
\cite{Roushan2017NP,Roushan2017Science,Xu2018PRL,Yan2019Science}, which thus
lacks one more dimension to realize the two-dimensional topological phenomena
induced by gauge potential, e.g., the quantum Hall effect or quantum spin Hall
effect%
%TCIMACRO{\TeXButton{TeX field}{~}}%
%BeginExpansion
~%
%EndExpansion
\cite{Shen2012Book}. Recently, the quasi-two-dimensional ladder model%
%TCIMACRO{\TeXButton{TeX field}{~}}%
%BeginExpansion
~%
%EndExpansion
\cite{Ye2019PRL}, and ture-two-dimensional Sycamore processor%
%TCIMACRO{\TeXButton{TeX field}{~}}%
%BeginExpansion
~%
%EndExpansion
\cite{Arute2019Nature} have both been achieved with the state-of-the-art
technology in superconducting quantum circuits, but neither of them involves
the research on artificial gauge potential. Therefore, the effect of
artificial gauge potential needs to be further explored beyond the
one-dimensional system. In particular, the ladder model is almost the
simpliest two-dimensional model that implies rich physics, which, for example,
can be mapped to the one-dimensional spin-orbit-coupled chain if\ penetrated
by the effective magnetic flux%
%TCIMACRO{\TeXButton{TeX field}{~}}%
%BeginExpansion
~%
%EndExpansion
\cite{Atala2014NP,Livi2016PRL}.

To make an initial attempt towards two-dimensional quantum simulation with
artificial gauge potential, we will design the concrete superconducting qubit
circuit that realizes the ladder model penetrated by the effective magnetic
flux. We will focus the vortex and Meissner phase transitions induced by
the\ competition of related parameters, such as the coupling strengths and
effective magnetic flux. Since the lattice number cannot be achieved so many
as the atom number in cold atom systems, we will mainly concentrate on the
practical case with finite lattice number. Besides, the method to measure the
two phases will also be discussed for the future experimental implementation.

In Sec.%
%TCIMACRO{\TeXButton{TeX field}{~}}%
%BeginExpansion
~%
%EndExpansion
\ref{sec:Model}, we introduce the theoretical model that employs two-tone
drives to engineer artificial gauge potential in the ladder model constructed
by superconducting qubit circuits. In Sec.%
%TCIMACRO{\TeXButton{TeX field}{~}}%
%BeginExpansion
~%
%EndExpansion
\ref{sec:phase transition}, we analyze the Vortex-Meissner phase transition at
different parameter regimes. In Sec.%
%TCIMACRO{\TeXButton{TeX field}{~}}%
%BeginExpansion
~%
%EndExpansion
\ref{Sec:ExpDetails}, we discuss the experimental feasibility to generate the
single-particle ground state and measure the vortex-Meissner phase transition.
In Sec.%
%TCIMACRO{\TeXButton{TeX field}{~}}%
%BeginExpansion
~%
%EndExpansion
\ref{sec:DC}, we summarize the results and make some discussions.

\section{Two-tone drive induced artificial gauge potential\label{sec:Model}}

\subsection{Theoretical model}

As an example, we consider the ladder model constructed by the X-shape
gradiometer flux qubit circuits (see schematic diagram in Fig.%
%TCIMACRO{\TeXButton{TeX field}{~}}%
%BeginExpansion
~%
%EndExpansion
\ref{fig:schematic diagram}). The individual flux qubit is manipulated by
classical direct current flux bias and alternating current drive (colored in
blue), and the states of qubits are dispersively read out through a coplanar
waveguide resonator (colored in green)%
%TCIMACRO{\TeXButton{TeX field}{~}}%
%BeginExpansion
~%
%EndExpansion
\cite{Blais2004PRA,Lin2014NC,Yan2016NC,Wu2018npj}. The flux qubits are coupled
to their nearest neighbours with mutual inductances that are mostly determined
by the nearest edge on the loop. The the flux qubit loop is designed like a
cross%
%TCIMACRO{\TeXButton{TeX field}{~}}%
%BeginExpansion
~%
%EndExpansion
\cite{Barends2013PRL} such that different coupling terms can be well separated
to minimize the crosstalk. The first (second) row of the qubits is called the
left (right) leg of the ladder.

The qubit parameters are assumed to be homogeneous along the leg. Then, the
bare Hamiltonian without driving fields can be generally given by%
\begin{align}
\hat{H}_{\text{b}}  &  =\sum_{d=\mathrm{L}}^{\mathrm{R}}\sum_{l}\frac{\hbar
}{2}\omega_{d}\hat{\sigma}_{z}^{\left(  d,l\right)  }\nonumber\\
&  -\sum_{d=\text{L,R}}\sum_{l}\hbar g_{0}\hat{\sigma}_{-}^{\left(
d,l\right)  }\hat{\sigma}_{+}^{\left(  d,l+1\right)  }+\text{H.c.},\nonumber\\
&  -\sum_{l}\hbar K_{0}\hat{\sigma}_{-}^{\left(  \text{L},l\right)  }%
\hat{\sigma}_{+}^{\left(  \text{R},l\right)  }+\text{H.c..} \label{eq:Hb}%
\end{align}
Here, according to the homogeneous assumption, all qubits along the $d$ leg
have the identical frequency $\omega_{d}$, where $d=\mathrm{L}$ or
$d=\mathrm{R}$ is the abbreviation of left or right. The bare intraleg
coupling strength on the left (L) or right (R) leg can be given by
$g_{d}=M_{d}I_{pd}^{2}/\hbar$ with $d=\mathrm{L,R}$\textrm{,} $M_{d}$ being
the mutual inductance between adjacent qubits (e.g., $\sim10%
%TCIMACRO{\TeXButton{TeX field}{\operatorname{pH}}}%
%BeginExpansion
\operatorname{pH}%
%EndExpansion
$), $I_{pd}$ the persistent current (e.g., $\sim0.1%
%TCIMACRO{\TeXButton{TeX field}{\operatorname{\mu A}}}%
%BeginExpansion
\operatorname{\mu A}%
%EndExpansion
$), and $\hbar$ the reduced Plank constant. The persistent current and the
qubit frequency can be tuned via designing the area ratio $\alpha$ between the
small and large junctions%
%TCIMACRO{\TeXButton{TeX field}{~}}%
%BeginExpansion
~%
%EndExpansion
\cite{Zhu2010APL}. Therefore, we can make the qubits on different legs of
distinct qubit frequencies. This also leads to $I_{p\text{L}}\neq
I_{p\text{R}}$, despite which, however, via careful design of $M_{d}$, we can
also make $g_{\text{L}}=g_{\text{R}}=g_{0}$ (e.g., $1\sim300%
%TCIMACRO{\TeXButton{TeX field}{\operatorname{MHz}\times2\pi}}%
%BeginExpansion
\operatorname{MHz}\times2\pi
%EndExpansion
$). Thus, in Eq.%
%TCIMACRO{\TeXButton{TeX field}{~}}%
%BeginExpansion
~%
%EndExpansion
(\ref{eq:Hb}), the intraleg coupling strengths on both legs are $g_{0}$.
Besides, $K_{0}$ denotes the interleg coupling strength, which is determined
by the interleg mutual inductance $M$ and also the persistent currents of the
flux qubit circuits on both legs, i.e., $K_{0}=M I_{p\mathrm{L}}
I_{p\mathrm{R}}$.

To engineer the effective magnetic flux from the bare Hamiltonian $\hat
{H}_{\text{b}}$, we will first show that the the qubit frequency can be
periodically modulated via the assist of classical driving fields, as will be
discussed below.

\subsection{Periodical modulation of the qubit frequency\label{sec:PM}}

We now demonstrate the periodical modulation of the qubit frequency through
two-tone drives. In our treatment, the flux qubit circuit is modelled as an
ideal two-level system because of the high anharmonicity%
%TCIMACRO{\TeXButton{TeX field}{~}}%
%BeginExpansion
~%
%EndExpansion
\cite{Orlandao1999PRB,Liu2005PRL,Robertson2006PRB} it possesses. In this
manner, the individual flux qubit at the $d$ leg and $l$th rung with two-tone
drives can be characterized by the Hamiltonian
\begin{equation}
\hat{H}_{d,l}%
%TCIMACRO{\TeXButton{TeX field}{\!}}%
%BeginExpansion
\!%
%EndExpansion
=%
%TCIMACRO{\TeXButton{TeX field}{\!}}%
%BeginExpansion
\!%
%EndExpansion
\frac{\hbar}{2}\omega_{d}\hat{\sigma}_{z}^{\left(  d,l\right)  }+\frac{\hbar
}{2}\sum_{j=1}^{2}\left[  \hat{\sigma}_{+}^{\left(  d,l\right)  }\Omega
_{j}^{\left(  d,l\right)  }e^{-i\omega_{j}^{\left(  d\right)  }t}%
+\mathrm{H.c.}\right]  , \label{eq:Hq}%
\end{equation}
in the qubit basis, where the $j$th driving field ($j=1,2$) possesses the
complex driving strength $\Omega_{j}^{\left(  d,l\right)  }$ at the frequency
$\omega_{j}^{\left(  d\right)  }$. However, the transmon qubit%
%TCIMACRO{\TeXButton{TeX field}{~}}%
%BeginExpansion
~%
%EndExpansion
\cite{Koch2007PRA,Sank2014Thesis} has a worse anharmonicity than the flux
qubit and thus, the detailed model should include the higher energy levels,
e.g., the second excited state (see Appendix.%
%TCIMACRO{\TeXButton{TeX field}{~}}%
%BeginExpansion
~%
%EndExpansion
\ref{appendix:PMQF}) .

In Eq.%
%TCIMACRO{\TeXButton{TeX field}{~}}%
%BeginExpansion
~%
%EndExpansion
(\ref{eq:Hq}), the driving field is determined by the incident current
$I_{j}^{\left(  d,l\right)  }%
%TCIMACRO{\TeXButton{TeX field}{\!}}%
%BeginExpansion
\!%
%EndExpansion
\left(  t\right)  $ through the relation $\operatorname{Re}\{\frac{\hbar}%
{2}\Omega_{j}^{\left(  d,l\right)  }e^{-i\omega_{j}^{\left(  d\right)  }%
t}\}=-M_{d}I_{pd}I_{j}^{\left(  d,l\right)  }%
%TCIMACRO{\TeXButton{TeX field}{\!}}%
%BeginExpansion
\!%
%EndExpansion
\left(  t\right)  $. The detunings of the driving frequencies $\omega
_{j}^{\left(  d\right)  }$ from the qubit frequencies $\omega_{d}$ are kept
identical for both ladder legs, i.e., $\delta_{j}\equiv\omega_{j}^{\left(
d\right)  }-\omega_{d}$ despite $d$ taking $\mathrm{L}$ or $\mathrm{R}$. In
fact, this can be achieved via tuning the driving frequencies $\omega
_{j}^{\left(  d\right)  }$ for the given qubit frequencies $\omega_{d}$.
Besides, we assume $\delta_{1}$ and $\delta_{2}$ are close to each other,
i.e., $\left\vert \delta\right\vert \ll|\delta_{1}|,|\delta_{2}|$ with
$\delta=\delta_{2}-\delta_{1}$. Also, we consider the large-detuning regime
$|\Omega_{j}^{\left(  d,l\right)  }/\delta_{j}|^{2}\ll1$ and homogeneous
(inhomogeneous) driving strengths (phases), i.e., $\Omega_{1}^{\left(
d,l\right)  }=\Omega_{1}$ and $\Omega_{2}^{\left(  d,l\right)  }=\Omega
_{2}e^{-i\phi_{d,l}}$ with positive $\Omega_{j}$. Then, via the second-order
perturbative method, the effective Hamiltonian can be yielded (see Appendix.%
%TCIMACRO{\TeXButton{TeX field}{~}}%
%BeginExpansion
~%
%EndExpansion
\ref{appendix:PMQF}) as
\begin{equation}
\hat{H}_{d,l}^{\left(  \text{eff}\right)  }%
%TCIMACRO{\TeXButton{TeX field}{\!}}%
%BeginExpansion
\!%
%EndExpansion
=%
%TCIMACRO{\TeXButton{TeX field}{\!}}%
%BeginExpansion
\!%
%EndExpansion
\frac{\hbar}{2}\omega_{d}\hat{\sigma}_{z}^{\left(  d,l\right)  }%
%TCIMACRO{\TeXButton{TeX field}{\!}}%
%BeginExpansion
\!%
%EndExpansion
-\frac{\hbar}{2}%
%TCIMACRO{\TeXButton{TeX field}{\!}}%
%BeginExpansion
\!%
%EndExpansion
\left[  \omega_{s}%
%TCIMACRO{\TeXButton{TeX field}{\!}}%
%BeginExpansion
\!%
%EndExpansion
+%
%TCIMACRO{\TeXButton{TeX field}{\!}}%
%BeginExpansion
\!%
%EndExpansion
\Omega\cos(\delta t%
%TCIMACRO{\TeXButton{TeX field}{\!}}%
%BeginExpansion
\!%
%EndExpansion
+%
%TCIMACRO{\TeXButton{TeX field}{\!}}%
%BeginExpansion
\!%
%EndExpansion
\phi^{\left(  d,l\right)  })\right]
%TCIMACRO{\TeXButton{TeX field}{\!}}%
%BeginExpansion
\!%
%EndExpansion
\hat{\sigma}_{z}^{\left(  d,l\right)  }, \label{ea:Hq_modulated}%
\end{equation}
where $\omega_{s}=\sum_{j=1}^{2}\frac{\Omega_{j}{}^{2}}{2\delta_{j}}$ is the
Stark shift and $\Omega=|\frac{\Omega_{1}\Omega_{2}}{\delta_{1}}|$. The phase
$\phi^{\left(  d,l\right)  }$ can be tuned by the driving field at the site
$(d,l)$, which will not be specified at present.

In Eq.%
%TCIMACRO{\TeXButton{TeX field}{~}}%
%BeginExpansion
~%
%EndExpansion
(\ref{ea:Hq_modulated}), we find that the qubit frequency is periodically
modulated with the strength $\Omega$, the frequency $\delta$, and the phase
$\phi^{\left(  d,l\right)  }$. Under our assumption, the parameters can be
typically, $\delta_{1}^{\left(  d\right)  }/2\pi=1~\mathrm{GHz}$\textrm{,
}$\delta_{2}^{\left(  d\right)  }/2\pi=1.1~\mathrm{GHz}$, and $\Omega_{1}%
/2\pi=\Omega_{2}/2\pi=178$~$\mathrm{MHz}$, in which case, the Stark shift
$\omega_{s}/2\pi=30.24~\mathrm{MHz}$, the modulation strength is $\Omega
/2\pi=31.7~\mathrm{MHz}$ and the modulation frequency $\delta/2\pi
=100~\mathrm{MHz}$. The qubit frequency $\omega_{d}/2\pi\ $can be about
$2~\mathrm{GHz}$, which, together with the driving frequencies $\omega
_{j}^{\left(  d\right)  }$, is left to be exactly determined in the following.

Note that one driving field will only arouse transitions between qubit bases
[see the individual driving term in Eq.%
%TCIMACRO{\TeXButton{TeX field}{~}}%
%BeginExpansion
~%
%EndExpansion
(\ref{eq:Hq})]. That's why we apply two-tone driving fields to achieve the
periodical modulation of the qubit frequency.

We must also mention that the method introduced here is applicable for all
qubit circuits, and not merely confined to the flux qubit (see Appendix.%
%TCIMACRO{\TeXButton{TeX field}{~}}%
%BeginExpansion
~%
%EndExpansion
\ref{appendix:PMQF}). Its validity does not require a negligibly small
anharmonicity of the qubit circuit as that for the transmon circuit in Ref.%
%TCIMACRO{\TeXButton{TeX field}{~}}%
%BeginExpansion
~%
%EndExpansion
\cite{Alaeian2019PRA}. Since nonlinearity is a key factor for demonstrating
quantum phenomena%
%TCIMACRO{\TeXButton{TeX field}{~}}%
%BeginExpansion
~%
%EndExpansion
\cite{Wendin2007LTP}, periodically modulating the qubit frequency while
maintaining enough anharmonicity can be significant for exploring
nonequilibrium quantum physics.

\subsection{Engineering effective magnetic flux}

Based on the periodical modulation of the qubit frequency in Sec.%
%TCIMACRO{\TeXButton{TeX field}{~}}%
%BeginExpansion
~%
%EndExpansion
\ref{sec:PM}, we now continue to demonstrate how to engineer the effective
magnetic flux. We assume each qubit in Fig.%
%TCIMACRO{\TeXButton{TeX field}{~}}%
%BeginExpansion
~%
%EndExpansion
\ref{fig:schematic diagram} is driven by two-tone fields such that the qubit
frequency can be modulated as in Eq.%
%TCIMACRO{\TeXButton{TeX field}{~}}%
%BeginExpansion
~%
%EndExpansion
(\ref{ea:Hq_modulated}). To include the nearest qubit-qubit couplings, the
full Hamiltonian can be represented as%
\begin{equation}
\hat{H}_{\text{f}}=\hat{H}_{\text{b}}-\sum_{d=\mathrm{L}}^{\mathrm{R}}\sum
_{l}\frac{\hbar}{2}\left[  \omega_{s}+\Omega\cos(\delta t+\phi^{\left(
d,l\right)  })\right]  \hat{\sigma}_{z}^{\left(  d,l\right)  }. \label{eq:H}%
\end{equation}
Note that $\hat{H}_{\text{b}}$ is the bare Hamiltonian given in Eq.%
%TCIMACRO{\TeXButton{TeX field}{~}}%
%BeginExpansion
~%
%EndExpansion
(\ref{eq:Hb}), $\omega_{s}$ is the Stark shift, and $\Omega$, $\delta$, and
$\phi^{(d,l)}$ are respectively the periodical modulation strength, frequency,
and phase of the qubit at $(d,l)$.

To eliminate the time-dependent terms in Eq.%
%TCIMACRO{\TeXButton{TeX field}{~}}%
%BeginExpansion
~%
%EndExpansion
(\ref{eq:H}), we now apply to Eq.%
%TCIMACRO{\TeXButton{TeX field}{~}}%
%BeginExpansion
~%
%EndExpansion
(\ref{eq:H}) a unitary transformation%
\begin{equation}
\hat{U}_{d}\left(  t\right)  =%
%TCIMACRO{\tprod \limits_{l}}%
%BeginExpansion
{\textstyle\prod\limits_{l}}
%EndExpansion%
%TCIMACRO{\tprod \limits_{d=\mathrm{L,R}}}%
%BeginExpansion
{\textstyle\prod\limits_{d=\mathrm{L,R}}}
%EndExpansion
\exp\left[  i\hat{F}_{l,d}\left(  t\right)  \right]  ,
\end{equation}
where the expression of $\hat{F}_{l,d}\left(  t\right)  $ is explicitly given
by
\begin{equation}
\hat{F}_{l,d}\left(  t\right)  =\frac{1}{2}\hat{\sigma}_{z}^{\left(
d,l\right)  }\left[  \frac{\Omega}{\delta}\sin\left(  \delta t+\phi
_{d,l}\right)  +\left(  \omega_{d}-\omega_{s}\right)  t\right]  .
\end{equation}
After that, the assumptions $\phi_{d\text{,}l}=\phi_{d}-\phi l$,
$\phi_{\text{L}}=-\phi_{\text{R}}=\phi_{0}$, and $\delta=\omega_{\text{R}%
}-\omega_{\text{L}}$ are made and the fast-oscillating terms are neglected,
thus leading to the following qubit ladder Hamiltonian as
\begin{align}
\hat{H}_{\text{f}}^{\prime}=  &  -\sum_{d=\text{L}}^{\text{R}}\sum_{l}\hbar
g\hat{\sigma}_{-}^{\left(  d,l\right)  }\hat{\sigma}_{+}^{\left(
d,l+1\right)  }+\text{H.c.}\nonumber\\
&  -\sum_{l}\hbar K\hat{\sigma}_{-}^{\left(  \text{L},l\right)  }\hat{\sigma
}_{+}^{\left(  \text{R},l\right)  }\exp\left(  i\phi l\right)  +\text{H.c..}
\label{eq:Hld0}%
\end{align}
Here, the intraleg coupling strength $g=g_{0}J_{0}\left(  \eta_{x}\right)  $,
and the interleg coupling strength $K=K_{0}J_{1}\left(  \eta_{y}\right)  $,
which are in principle tunable via modifying $\Omega$, since $\eta_{x}%
=\frac{2\Omega}{\delta}\sin(\frac{\phi}{2})$ and $\eta_{y}=\frac{2\Omega
}{\delta}\sin\left(  \phi_{0}\right)  $ (see Appendix.%
%TCIMACRO{\TeXButton{TeX field}{~}}%
%BeginExpansion
~%
%EndExpansion
\ref{append:Treatment} for details). The symbol $J_{n}\left(  x\right)  $
represents the $n$th Bessel function of the first kind.

\begin{figure}[ptb]
\includegraphics[width=0.32\textwidth, clip]{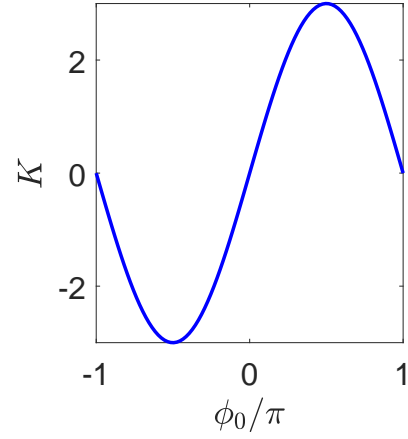}\caption{(color online).
Tunable interleg coupling strength $K$ plotted versus the phase $\phi_{0}$:
$K=3g\sin\phi_{0}$, where, for simplicity, we have set the intraleg coupling
strength $g=1$. Here, $\phi_{0}$ is determined by the phases of the driving
fields. }%
\label{fig:K_phi0}%
\end{figure}

For the typical parameters given previously, which yields $\Omega/2\pi=31.7%
%TCIMACRO{\unit{MHz}}%
%BeginExpansion
\operatorname{MHz}%
%EndExpansion
$ and $\delta/2\pi=100%
%TCIMACRO{\unit{MHz}}%
%BeginExpansion
\operatorname{MHz}%
%EndExpansion
$, we can further set $g_{0}/2\pi=3.5%
%TCIMACRO{\unit{MHz}}%
%BeginExpansion
\operatorname{MHz}%
%EndExpansion
$, and $K_{0}/2\pi=33%
%TCIMACRO{\unit{MHz}}%
%BeginExpansion
\operatorname{MHz}%
%EndExpansion
$. Then, the condition $|\eta_{x/y}/2|^{2}\ll1$ is fulfilled, which makes
$g\approx g_{0}$ and $K\approx\frac{\eta_{y}}{2}K_{0}=\frac{\Omega}{\delta
}K_{0}\sin\left(  \phi_{0}\right)  $. In this case, the intraleg coupling
strength is fixed at $g_{0}$, but the interleg coupling strength can also be
equivalently represented as%
\begin{equation}
K\approx3g\sin\phi_{0}.
\end{equation}
This implies that for given $g$, $K$ can be tuned via $\phi_{0}$ in the range
$-3g\leq K\leq3g$ (see Fig.%
%TCIMACRO{\TeXButton{TeX field}{~}}%
%BeginExpansion
~%
%EndExpansion
\ref{fig:K_phi0}), which enables us to study the phase transition by adjusting
$K$. The condition $\delta=\omega_{\text{R}}-\omega_{\text{L}}$ can be
satisfied with making the qubit frequencies $\omega_{\text{L}}/2\pi=1.9%
%TCIMACRO{\unit{GHz}}%
%BeginExpansion
\operatorname{GHz}%
%EndExpansion
$ and $\omega_{\text{R}}/2\pi=2%
%TCIMACRO{\unit{GHz}}%
%BeginExpansion
\operatorname{GHz}%
%EndExpansion
$ such that $\delta/2\pi=100%
%TCIMACRO{\unit{MHz}}%
%BeginExpansion
\operatorname{MHz}%
%EndExpansion
$. Furthermore, the driving frequencies should be $\omega_{1}^{\left(
\text{L}\right)  }/2\pi=2.9%
%TCIMACRO{\unit{GHz}}%
%BeginExpansion
\operatorname{GHz}%
%EndExpansion
$, $\omega_{2}^{\left(  \text{L}\right)  }/2\pi=3%
%TCIMACRO{\unit{GHz}}%
%BeginExpansion
\operatorname{GHz}%
%EndExpansion
$, $\omega_{1}^{\left(  \text{R}\right)  }/2\pi=3%
%TCIMACRO{\unit{GHz}}%
%BeginExpansion
\operatorname{GHz}%
%EndExpansion
$, and $\omega_{2}^{\left(  \text{R}\right)  }=3.1%
%TCIMACRO{\unit{GHz}}%
%BeginExpansion
\operatorname{GHz}%
%EndExpansion
$, since we have assumed $\delta_{1}/2\pi=(\omega_{1}^{\left(  d\right)
}-\omega_{d})/2\pi=1%
%TCIMACRO{\unit{GHz}}%
%BeginExpansion
\operatorname{GHz}%
%EndExpansion
$ and $\delta_{2}/2\pi=(\omega_{2}^{\left(  d\right)  }-\omega_{d})/2\pi=1.1%
%TCIMACRO{\unit{GHz}}%
%BeginExpansion
\operatorname{GHz}%
%EndExpansion
$.

So far, we have determined nearly all the necessary parameters of the qubit
and driving fields, except for the phases in the driving fields $\phi$ and
$\phi_{0}$, among which, the former acts as the effective magnetic flux per
plaquette, while the latter is used to tune the interleg coupling strength $K$.

\subsection{Fermionic ladder in the effective magnetic flux}

To transform the qubit ladder into the fermionic ladder, we can make a
Jordan-Wigner transformation%
%TCIMACRO{\TeXButton{TeX field}{~}}%
%BeginExpansion
~%
%EndExpansion
\cite{Mei2013PRB}, which is of the form as
\begin{align}
\hat{\sigma}_{-}^{\left(  \text{L},l\right)  }  &  =\hat{b}_{\text{L},l}%
\prod_{l^{\prime}=1}^{l-1}\exp%
%TCIMACRO{\TeXButton{TeX field}{\!}}%
%BeginExpansion
\!%
%EndExpansion
(i\pi\hat{b}_{\text{L},l}^{\dag}\hat{b}_{\text{L},l}%
),\label{eq:Jordan-Wigner1}\\
\hat{\sigma}_{-}^{\left(  \text{R},l\right)  }  &  =\hat{b}_{\text{R},l}%
\prod_{l^{\prime}=1}^{l}\exp(i\pi\hat{b}_{\text{L},l}^{\dag}\hat{b}%
_{\text{L},l})%
%TCIMACRO{\TeXButton{TeX field}{\!}}%
%BeginExpansion
\!%
%EndExpansion
\prod_{l^{\prime}=1}^{l-1}\exp(i\pi\hat{b}_{\text{R},l}^{\dag}\hat
{b}_{\text{R},l}). \label{eq:Jordan-Wigner2}%
\end{align}
Here, $\hat{\sigma}_{z}^{\left(  d,l\right)  }=2\hat{b}_{d,l}^{\dag}\hat
{b}_{d,l}-1$, and the fermionic anticommutation relations $\{\hat{b}%
_{d,l},\hat{b}_{d^{\prime},l^{\prime}}^{\dag}\}=\delta_{dd^{\prime}}%
\delta_{ll^{\prime}}$ and $\{\hat{b}_{d,l},\hat{b}_{d^{\prime},l^{\prime}%
}\}=0$ are fulfilled, where $\delta_{dd^{\prime}}$ and $\delta_{ll^{\prime}}$
are Kronecker delta functions. Then, the qubit ladder Hamiltonian $\hat
{H}_{\text{f}}^{\prime}$ in Eq.%
%TCIMACRO{\TeXButton{TeX field}{~}}%
%BeginExpansion
~%
%EndExpansion
(\ref{eq:Hld0}) can be transformed into the Hamiltonian of the fermionic
ladder, i.e.,
\begin{align}
\hat{H}_{\text{ld}}=  &  -\sum_{d=\text{L}}^{\text{R}}\sum_{l}\hbar g\hat
{b}_{d,l}\hat{b}_{d,l+1}^{\dag}+\text{H.c.}\nonumber\\
&  -\sum_{l}\hbar K\hat{b}_{\text{L},l}\hat{b}_{\text{R},l}^{\dag}\exp\left(
i\phi l\right)  +\text{H.c.,} \label{eq:Hld}%
\end{align}
which describes the motion of \textquotedblleft fermionic\textquotedblright%
\ particles, governed by the effective magnetic flux $\phi$. We note that the
above fermionic ladder model with effective magnetic flux can also be
interpreted as one-dimensional spin-orbit-coupled model%
%TCIMACRO{\TeXButton{TeX field}{~}}%
%BeginExpansion
~%
%EndExpansion
\cite{Atala2014NP,Livi2016PRL}, which may thus inspire the research towards
the realization of quantum spin Hall effect%
%TCIMACRO{\TeXButton{TeX field}{~}}%
%BeginExpansion
~%
%EndExpansion
\cite{Shen2012Book}. 

\section{Vortex-Meissner phase transition\label{sec:phase transition}}

\subsection{Infinite-length ladder\label{sec:InfLenLadder}}

\begin{figure}[ptb]
\includegraphics[width=0.52\textwidth, clip]{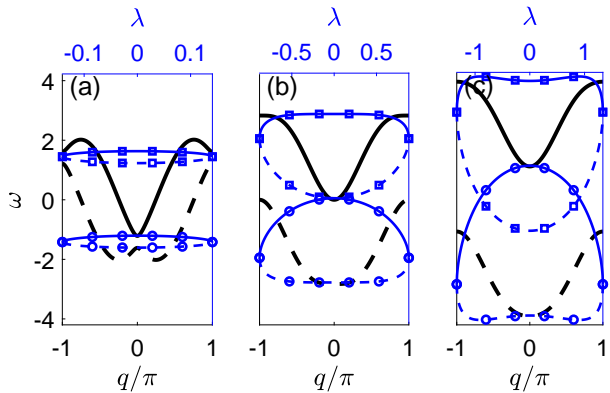}\caption{(color online)
Single-photon spectrum of the ladder model at the interleg coupling strength
$K=$ (a) $0.2$, (b) $\sqrt{2}$, and (c) $2.5$, respectively. Here, the
effective magnetic flux $\phi=\frac{\pi}{2}$, and the intraleg coupling
strength is set as unity: $g=1$, implying the critical interleg coupling
strength $K_{\mathrm{c}}=\sqrt{2}$. The solid (dashed) dark blue curve means
$\omega=\omega_{+}$ $(\omega_{-})$, and $z=\exp(iq)$; the solid (dashed) light
blue curve marked with \textquotedblleft$\square$\textquotedblright\ means
$\omega=\omega_{+}$ $(\omega_{-})$, and $z=\exp(\lambda)$; the solid (dashed)
light blue curve marked with \textquotedblleft$\circ$\textquotedblright\ means
$\omega=\omega_{+}$ $(\omega_{-})$, and $z=-\exp(\lambda)$. }%
\label{fig:spectrum}%
\end{figure}

Now, we seek the energy spectrum of the ladder Hamiltonian $\hat{H}%
_{\text{ld}}$ in the infinite chain case [see Eq.%
%TCIMACRO{\TeXButton{TeX field}{~}}%
%BeginExpansion
~%
%EndExpansion
(\ref{eq:Hld})], i.e., the lattice site (or rung) number $N$ approaches
infinity. To do this, we straightforwardly assume that the single-particle
eigenstate at the energy $\hbar\omega$\ is $\left\vert \omega\right\rangle
=\sum_{d,l}\psi_{d,l}\left\vert d,l\right\rangle $. Here, the notation
$\left\vert d,l\right\rangle =\hat{b}_{d,l}^{\dag}\left\vert 0\right\rangle $
represents the single-particle state at the site $\left(  \mathrm{L},l\right)
$ and $\left\vert 0\right\rangle $ is the ground state. Afterwards, we assume
the wave function $\psi_{d,l}\equiv\psi_{d,l}\left(  z\right)  $, and further
assume
\begin{equation}
\psi_{\text{L},l}=\psi_{\text{L},0}z^{l}e^{-i\frac{\phi}{2}l}\text{ and }%
\psi_{\text{R,}l}=\psi_{\text{R,}0}z^{l}e^{i\frac{\phi}{2}l}
\label{eq:psi_d_l_z}%
\end{equation}
before substituting the eigenstate vector $\left\vert \omega\right\rangle $
into the following secular equation
\begin{equation}
\hat{H}_{\text{ld}}\left\vert \omega\right\rangle =\hbar\omega\left\vert
\omega\right\rangle .
\end{equation}
Then, the dispersion relation can be yielded as a two-band spectrum i.e.,
\begin{equation}
\omega=\omega_{\pm}=-2gz_{\text{p}}^{2}\cos\frac{\phi}{2}\pm\sqrt{K^{2}%
-4g^{2}z_{\text{m}}^{2}\sin^{2}\frac{\phi}{2}}, \label{eq:dispersion}%
\end{equation}
where the intermediate parameters $z_{\text{p}}=(z+ z^{-1})/2$ and
$z_{\text{m}}=(z- z^{-1})/2$. The corresponding wave function at $l=0$ is of
the form
\begin{equation}
\psi_{\text{L},0}%
%TCIMACRO{\TeXButton{TeX field}{\!}}%
%BeginExpansion
\!%
%EndExpansion
\left(  z\right)  =\omega+g(ze^{i\frac{\phi}{2}}+z^{-1}e^{-i\frac{\phi}{2}%
})\text{ and }\psi_{\text{R,}0}%
%TCIMACRO{\TeXButton{TeX field}{\!}}%
%BeginExpansion
\!%
%EndExpansion
\left(  z\right)  =-K,
\end{equation}
where a global normalized constant has been discarded.

To guarantee the existence of $\omega$, there can be the following three
cases, i.e., (i) $z=\exp\left(  iq\right)  $, (ii) $z=\exp\left(
\lambda\right)  $, and (iii) $z=-\exp\left(  \lambda\right)  $, where $q$ and
$\lambda$ must be in the regime $-\pi\leq q\leq\pi$ and $-\ln\Lambda
\leq\lambda\leq\ln\Lambda$, where the parameter $\Lambda=K/2g\sin\frac{\phi
}{2}+\sqrt{K^{2}/4g^{2}\sin^{2}\frac{\phi}{2}+1}$. Here, the case (i) gives a
transmission state, the case (ii) a decay state, and the case (iii) a
staggered decay state. In the case (i), the value of $K$ can control the
number of the minimums of $\omega_{-}$, for which, there exists a critical
interleg coupling strength with the analytical form
\begin{equation}
K_{\text{c}}=2g\tan\frac{\phi}{2}\sin\frac{\phi}{2}. \label{eq:Kc}%
\end{equation}
The relation $K=K_{\text{c}}$ actually yields the vortex-Meissner transition
boundary discussed afterwards. In detail, if $K<$ $K_{\text{c}}$, the lower
band $\omega_{-}$ has two minimums, while, otherwise, the minimum number is
one. This can be clearly found from the dashed black curve in Fig.%
%TCIMACRO{\TeXButton{TeX field}{~}}%
%BeginExpansion
~%
%EndExpansion
\ref{fig:spectrum}(a)-\ref{fig:spectrum}(c) for $K$ taking $0.5$, $\sqrt{2}$,
and $2.5$, respectively, where we specify $g=1$ and $\phi=\frac{\pi}{2}$ such
that $K_{\text{c}}=\sqrt{2}$. As $K$ is increased, the band gap between the
two transmission bands $\omega_{+}$ and $\omega_{-}$ will also be broadened.
In Fig.%
%TCIMACRO{\TeXButton{TeX field}{~}}%
%BeginExpansion
~%
%EndExpansion
\ref{fig:spectrum}, where the energy bands $\omega_{\pm}$ for the decay and
staggered decay states have also been shown, we also find that a given
single-particle energy will always correspond to four degenerate states. This
is critical for the existence of the single-particle eigenstates under the
open boundary condition, which can in principle be constructed by the linear
superposition of these four degenerate states. Only when the decay and
staggered decay states are included, one can definitely ensure the equality
between the number of the independent coefficients and that of the boundary
conditions, considering that there are four terminals of the ladder. However,
in the simplest one-dimensional chain, which has only two terminals, the
single-particle eigenstates under the open boundary condition is only the
superposition of two transmission states, which differs from the
quasi-two-dimensonal ladder model this present paper concentrates on.

\subsection{Open-boundary ladder with finite qubit number}

Now we invstigate the open-boundary condition for the ladder model. In cold
atom systems, the ideal open-boundary effect is a hard wall, which is very
hard to realize
%TCIMACRO{\TeXButton{TeX field}{~}}%
%BeginExpansion
~%
%EndExpansion
\cite{Atala2014Thesis}, and the open-boundary condition is approximately
engineered by an external power law potential. However, in superconducting
qubit systems, the open-boundary condition is very convenient to realize,
since the ladder length is finite in experiment. Suppose the ladder length is
$N$, then the fermionic Hamiltonian in Eq.%
%TCIMACRO{\TeXButton{TeX field}{~}}%
%BeginExpansion
~%
%EndExpansion
(\ref{eq:Hld}) becomes
\begin{align}
\hat{H}_{\text{ld}}^{\left(  N\right)  }=  &  -\sum_{l=1}^{N-1}\sum
_{d=\text{L}}^{\text{R}}\hbar g\hat{b}_{d,l}\hat{b}_{d,l+1}^{\dag}%
+\text{H.c.}\nonumber\\
&  -\sum_{l=1}^{N}\hbar K\hat{b}_{\text{L},l}\hat{b}_{\text{R},l}^{\dag}%
\exp\left(  i\phi l\right)  +\text{H.c.,} \label{eq:Hld_N}%
\end{align}
where the eigenstates are different from those of the infinite-length ladder,
and therefore must be revisited. In Figs.%
%TCIMACRO{\TeXButton{TeX field}{~}}%
%BeginExpansion
~%
%EndExpansion
\ref{fig:spectrum}(a)-\ref{fig:spectrum}(c), we find that in infinite-length
case, a definite $\omega$ corresponds to four states, which we denote by the
characteristic constants $z=z_{1},$ $z_{2},$ $z_{3},$ and $z_{4}$,
respectively. In our study, we are only interested in the low-energy states.
Thus, the parameters $z_{j}$ can be determined by the relation $\omega
=\omega_{-}$ [see Eq.%
%TCIMACRO{\TeXButton{TeX field}{~}}%
%BeginExpansion
~%
%EndExpansion
(\ref{eq:dispersion})], which yields
\begin{align}
z_{1,2}  &  \equiv z_{1,2}\left(  \omega\right)  =\frac{1}{2}(R_{-}\mp
\sqrt{R_{-}^{2}-4}),\label{eq:z12}\\
z_{3,4}  &  \equiv z_{3,4}\left(  \omega\right)  =\frac{1}{2}(R_{+}\mp
\sqrt{R_{+}^{2}-4}), \label{eq:z34}%
\end{align}
with the compact symbols $R_{\pm}$, determined by $\omega$, represented in the
form as
\begin{equation}
R_{\pm}=-\frac{\omega}{g}\cos\frac{\phi}{2}\pm\sqrt{-\frac{\omega^{2}}{g^{2}%
}\sin^{2}\frac{\phi}{2}+\frac{K^{2}}{g^{2}}+4\sin^{2}\frac{\phi}{2}}.
\end{equation}

\begin{figure}[ptb]
\includegraphics[width=0.54\textwidth, clip]{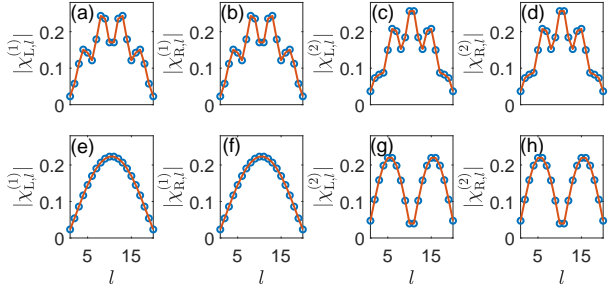}\caption{(color online).
Probability amplitude $|\chi_{d,l}^{\left(  n\right)  }|$ for the lowest two
states $\chi_{d,l}^{\left(  n\right)  }$ ($n=1,2$, and $d=$L,R) for the energy
$\mu_{n}$ in the open-boundary condition. Here, $n$ denotes the index of the
energy level, $d$ the ladder leg, and $l$ the rung index. The
\textquotedblleft$\circ$\textquotedblright\ marks the direct numerical
diagonalization result, and the solid curve is the fitted result using the
expansion equation $\chi_{d,l}^{\left(  n\right)  }=\sum_{j=1}^{4}%
A_{j}^{\left(  n\right)  }\chi_{d,l}^{\left(  n,j\right)  }$, where
$\chi_{d,l}^{\left(  n,j\right)  }$ is the $j$th transmission or decay state
in the infinite-length condition for the energy $\hbar\mu_{n}$. In (a)-(d),
the interleg coupling strength $K=0.5$, while in (e)-(h), $K=2.5$. The
intraleg coupling strength $g=1$, the effective magnetic flux per plaquette
$\phi=\frac{\pi}{2}$ for which $K_{\text{c}}=\sqrt{2}$, and the ladder length
$N=20$.}%
\label{fig:wavefunction}%
\end{figure}

For the open-boundary ladder with finite qubit number, the single-particle
eigenstate at the energy $\hbar\mu$ can be assumed as
\begin{equation}
\left\vert \mu\right\rangle =\sum_{d=\text{L}}^{\text{R}}\sum_{l=1}^{N}%
\chi_{d,l}\left\vert d,l\right\rangle . \label{eq:mu_expansion}%
\end{equation}
Here, the eigen wave function $\chi_{d,l}\equiv\chi_{d,l}%
%TCIMACRO{\TeXButton{TeX field}{\!}}%
%BeginExpansion
\!%
%EndExpansion
\left(  \mu\right)  $ must be the linear superposition of the four degenerate
states at the energy $\omega=\mu$ of the infinite-length ladder, respectively
denoted as $\psi_{d,l}^{\left(  j\right)  }\equiv\psi_{d,l}\left(
z_{j}\left(  \mu\right)  \right)  $ [see Eqs.%
%TCIMACRO{\TeXButton{TeX field}{~}}%
%BeginExpansion
~%
%EndExpansion
(\ref{eq:psi_d_l_z}), (\ref{eq:z12}), and (\ref{eq:z34})], i.e.,
\begin{equation}
\chi_{d,l}=\sum_{j=1}^{4}%
%TCIMACRO{\TeXButton{TeX field}{\!}}%
%BeginExpansion
\!%
%EndExpansion
A_{j}\psi_{d,l}^{\left(  j\right)  }.
\end{equation}
Then, by substituting the state vector expansion $\left\vert \mu\right\rangle
$ in\ Eq.%
%TCIMACRO{\TeXButton{TeX field}{~}}%
%BeginExpansion
~%
%EndExpansion
(\ref{eq:mu_expansion}) into the secular equation%
\begin{equation}
\hat{H}_{\text{ld}}^{\left(  N\right)  }\left\vert \mu\right\rangle =\hbar
\mu\left\vert \mu\right\rangle , \label{eq:secular_eq}%
\end{equation}
where the coefficients $A_{j}$ must be constrained nonzero, the eigen energies
can in principle be discretized as $\mu=\hbar\mu_{n}$ ($n=1,2,...,2N$) with
$\mu_{n}\leq\mu_{n+1}$ and the corresponding eigenstates can be assumed of the
form
\begin{equation}
\left\vert \mu_{n}\right\rangle =\sum_{d=\text{L}}^{\text{R}}\sum_{l=1}%
^{N}\chi_{d,l}^{\left(  n\right)  }%
%TCIMACRO{\TeXButton{TeX field}{\!}}%
%BeginExpansion
\!%
%EndExpansion
\left\vert d,l\right\rangle .
\end{equation}
Here, the lowest energy eigenstate $|\mu_{1}\rangle$ is called the
single-particle ground state, which is the major state we will study. The
eigen wave function $\chi_{d,l}^{\left(  n\right)  }$ can also be expanded as
the linear superposition of $\psi_{d,l}^{\left(  n,j\right)  }\equiv\psi
_{d,l}\left(  z_{j}\left(  \mu_{n}\right)  \right)  $, the degenerate states
in the infinite-length case, i.e.,
\begin{equation}
\chi_{d,l}^{\left(  n\right)  }=\sum_{j=1}^{4}A_{j}^{\left(  n\right)  }%
\psi_{d,l}^{\left(  n,j\right)  }. \label{eq:expand_inf}%
\end{equation}
However, straightforwardly solving Eq.%
%TCIMACRO{\TeXButton{TeX field}{~}}%
%BeginExpansion
~%
%EndExpansion
(\ref{eq:secular_eq}) is difficult, since a transcendental equation will be
involved. Thus, in this paper, the determination of $A_{j}^{\left(  n\right)
}$ is achieved by fitting Eq.%
%TCIMACRO{\TeXButton{TeX field}{~}}%
%BeginExpansion
~%
%EndExpansion
(\ref{eq:expand_inf}) with the results obtained from direct numerical
diagonalization of Eq.%
%TCIMACRO{\TeXButton{TeX field}{~}}%
%BeginExpansion
~%
%EndExpansion
(\ref{eq:secular_eq}).

In Fig.%
%TCIMACRO{\TeXButton{TeX field}{~}}%
%BeginExpansion
~%
%EndExpansion
\ref{fig:wavefunction}, the wave functions of the single-particle ground state
$\left\vert \mu_{1}\right\rangle $ and single-particle excited state
$\left\vert \mu_{2}\right\rangle $ ($\mu_{1}<\mu_{2}$) have been shown for
$K\ $taking $0.5$ [see Fig.%
%TCIMACRO{\TeXButton{TeX field}{~}}%
%BeginExpansion
~%
%EndExpansion
\ref{fig:wavefunction}(a)-\ref{fig:wavefunction}(d)] and $2.5$ [see Fig.%
%TCIMACRO{\TeXButton{TeX field}{~}}%
%BeginExpansion
~%
%EndExpansion
\ref{fig:wavefunction}(e)-\ref{fig:wavefunction}(h)], respectively, where the
other parameters are $g=1$, $N=20$, and $\phi=\frac{\pi}{2}$ such that
$K_{\text{c}}=\sqrt{2}$. The discrete circles represent the results from the
direct numerical diagonalization using Eq.%
%TCIMACRO{\TeXButton{TeX field}{~}}%
%BeginExpansion
~%
%EndExpansion
(\ref{eq:secular_eq}), while the solid curves the fitting results using the
expansion equation in Eq.%
%TCIMACRO{\TeXButton{TeX field}{~}}%
%BeginExpansion
~%
%EndExpansion
(\ref{eq:expand_inf}). Both results can be found to fit each other exactly.
Also, the wave functions at $K=2.5>K_{\text{c}}$ appear smoother than those at
$K=0.5<K_{\text{c}}$. Besides, when $K=2.5$, $|\chi_{d,l}^{\left(  2\right)
}|$ exhibits an obvious dip near the middle lattice site, which nevertheless
does not occur when $K=0.5$.

Then, we investigate the properties of the single-particle ground state
$\chi_{d,l}^{\left(  1\right)  }$ using the expansion coefficients
$A_{j}^{\left(  n\right)  }$ from fitting. From the discussions in Sec.%
%TCIMACRO{\TeXButton{TeX field}{~}}%
%BeginExpansion
~%
%EndExpansion
\ref{sec:InfLenLadder}, we know that if $K\ $is less than $K_{\text{c}}$, all
the four characteristic constants $z_{j}$ correpsonding to $\omega=\mu_{1}$
are complex numbers on the unit circle, while, if $K$ exceeds $K_{\text{c}}$,
$z_{3}$ and $z_{4}$ will become real, which will only contribute to the
population at the edges. Due to the effective magnetic flux, a complex
characteristic constant $z_{j}=\exp\left(  iq_{j}\right)  $ corresponds to a
plane wave with the quasimomentum $q_{j}-\phi/2$ ($q_{j}+\phi/2$) in the wave
function of the L (R) ladder leg [see Eq.~(\ref{eq:psi_d_l_z})].

\begin{figure}[ptb]
\includegraphics[width=0.42\textwidth, clip]{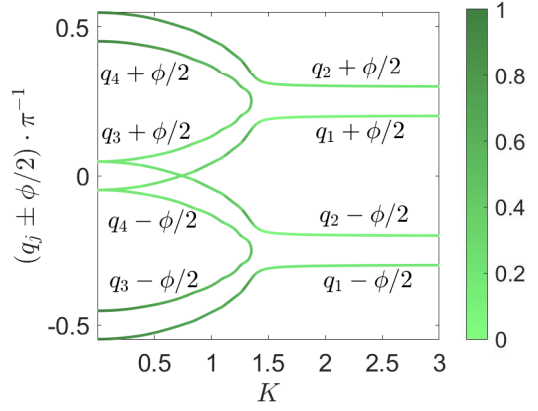}\caption{(color online).
Quasimomentum $q_{j}\pm\phi/2$ in the single-particle ground state wave
function $\chi_{d,l}^{\left(  1\right)  }$ for different interleg coupling
strength $K$. Here, the effective magnetic flux $\phi=\pi/2$, the intraleg
coupling strength $g=1$, and the ladder length $N=20$. The color indicates the
relative distribution intensity of the wave function on the quasimomentum
component. Here, the quasimomentum $q_{j}-\phi/2$ ($q_{j}+\phi/2$) only occurs
on the L (R) ladder leg. }%
\label{fig:ComponentSpectrum}%
\end{figure}

In Fig.%
%TCIMACRO{\TeXButton{TeX field}{~}}%
%BeginExpansion
~%
%EndExpansion
\ref{fig:ComponentSpectrum}, we have plotted the quasimomentum $q_{j}\mp
\phi/2$ versus the interleg coupling strength $K$ with $\phi=\pi/2$ and
$N=20$,\ where the color represents the relative distribution intensity on a
particular quasimomentum component [obtained by rescaling $|A_{j}^{\left(
1\right)  }\psi_{d,0}^{\left(  1,j\right)  }|$, with $d$ taking L (R) for
$q_{j}-\phi/2$ ($q_{j}+\phi/2$)]. We can also see that if $\phi=\pi/2$, and
$K$ is less than $K_{\text{c}}$, the particle is more likely to be populated
on the L (R) leg, corresponding to the characteristic constant $z_{1,3}$
($z_{2,4}$). However, if $K$ exceeds $K_{\text{c}}$, only $z_{1}$ and $z_{2}$
remain complex, and the particles corresponding to $z_{1,2}$ are approximately
populated uniformly on both legs.

Lastly, we mention that once the single-particle eigenstates $\chi
_{d,l}^{\left(  n\right)  }$ are obtained, one can make the transformation
$\hat{b}_{n}^{\dag}=\sum_{d=\text{L}}^{\text{R}}\sum_{l=1}^{N}\chi
_{d,l}^{\left(  n\right)  }\hat{b}_{d,l}^{\dag}$, which can finally transform
the Hamiltonian in Eq.%
%TCIMACRO{\TeXButton{TeX field}{~}}%
%BeginExpansion
~%
%EndExpansion
(\ref{eq:Hld_N}) into the independent fermionic modes, i.e.,
\begin{equation}
\hat{H}_{\text{ld}}^{\left(  N\right)  }=\sum_{n=1}^{2N}\hbar\mu_{n}\hat
{b}_{n}^{\dag}\hat{b}_{n}.
\end{equation}
Here, $\hat{b}_{n}$ and $\hat{b}_{n}^{\dag}$ meet the fermionic
anticommutation relations, i.e., $\{\hat{b}_{n},\hat{b}_{n^{\prime}}^{\dag
}\}=\delta_{nn^{\prime}}$. Compared with the infinite-length scenario, we note
that the eigen energies are discretized, with the eigenstates being the
superposition of the ones in the infinite-length scenario.

\subsection{Chiral current}

The current operator can be derived from the following continuity equation
\begin{equation}
\frac{\text{d}}{\text{d}t}(\hat{b}_{d,l}^{\dag}\hat{b}_{d,l})=\frac{[\hat
{b}_{d,l}^{\dag}\hat{b}_{d,l},\hat{H}_{\text{ld}}]}{i\hbar}=%
%TCIMACRO{\TeXButton{hat_{j}}{\hat{j}}}%
%BeginExpansion
\hat{j}%
%EndExpansion
_{l-1,l}^{\left(  d\right)  }+%
%TCIMACRO{\TeXButton{hat_{j}}{\hat{j}}}%
%BeginExpansion
\hat{j}%
%EndExpansion
_{l+1,l}^{\left(  d\right)  }+%
%TCIMACRO{\TeXButton{hat_{j}}{\hat{j}}}%
%BeginExpansion
\hat{j}%
%EndExpansion
_{l,\bar{d}d},
\end{equation}
where $d,\bar{d}\in\{$L,R$\}$ and $\bar{d}\neq d$. Here, $%
%TCIMACRO{\TeXButton{hat_{j}}{\hat{j}}}%
%BeginExpansion
\hat{j}%
%EndExpansion
_{l,l+1}^{\left(  d\right)  }$ denotes the particle current flowing from the
site $l$ to $l+1$ on the $d$ ladder, while $%
%TCIMACRO{\TeXButton{hat_{j}}{\hat{j}}}%
%BeginExpansion
\hat{j}%
%EndExpansion
_{l,\bar{d}d}$ the particle current flowing from the $\bar{d}$ ladder to $d$
ladder at the $l$th site. The physical meaning is that the time-varying rate
of the particle number at one individual site is determined by the current
that flows into it. The resulting current operator can be explicitly
represented as
\begin{align}%
%TCIMACRO{\TeXButton{hat_{j}}{\hat{j}}}%
%BeginExpansion
\hat{j}%
%EndExpansion
_{l,l+1}^{\left(  d\right)  }  &  =ig(\hat{b}_{d,l+1}^{\dag}\hat{b}_{d,l}%
-\hat{b}_{d,l}^{\dag}\hat{b}_{d,l+1}),\\%
%TCIMACRO{\TeXButton{hat_{j}}{\hat{j}}}%
%BeginExpansion
\hat{j}%
%EndExpansion
_{l,\text{LR}}  &  =iK(\hat{b}_{\text{R},l}^{\dag}\hat{b}_{\text{L},l}e^{i\phi
l}-\hat{b}_{\text{L},l}^{\dag}\hat{b}_{\text{R},l}e^{-i\phi l}).
\end{align}
For the specific single-particle ground state $\left\vert \mu_{1}\right\rangle
=\sum_{d=\text{L}}^{\text{R}}\sum_{l=1}^{N}\chi_{d,l}^{\left(  1\right)
}\left\vert d\text{,}l\right\rangle $, the average particle current can be
respectively given by
\begin{equation}
j_{l,l+1}^{\left(  d\right)  }=ig(\chi_{d,l+1}^{\left(  1\right)  \ast}%
\chi_{d,l}^{\left(  1\right)  }-\chi_{d,l+1}^{\left(  1\right)  }\chi
_{d,l}^{\left(  1\right)  \ast})
\end{equation}
which describes the flow from the site $l$ to $l+1$ on the $d$ ladder, and
\begin{equation}
j_{l,\text{LR}}=iK(\chi_{\text{R},l}^{\left(  1\right)  \ast}\chi_{\text{L}%
,l}^{\left(  1\right)  }e^{i\phi l}-\chi_{\text{R},l}^{\left(  1\right)  }%
\chi_{\text{L},l}^{\left(  1\right)  \ast}e^{-i\phi l})
\end{equation}
which describes the flow from the L to R ladder at the $l$th site.

\begin{figure}[ptb]
\includegraphics[bb=19 8 430 443, width=0.40\textwidth, clip]{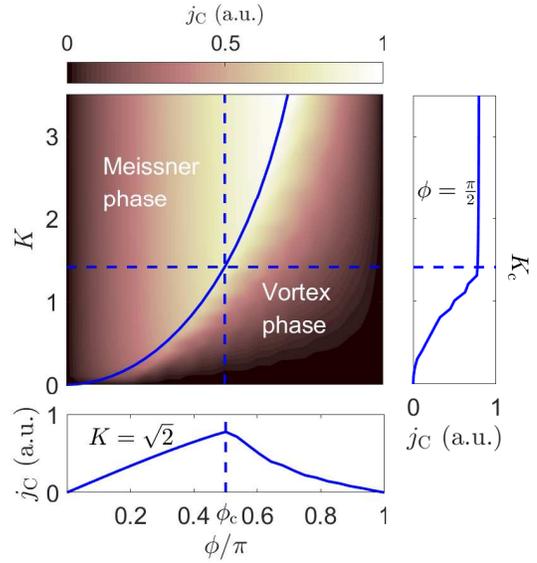}\caption{(color
online). Chiral current strengths $j_{\text{C}}$ as a function of the
effective magnetic flux $\phi$ and the interleg coupling $K$ with $N=20$
sites, $g=1$, and open boundary conditions. The solid curve is the critical
boundary separating the Meissner and vortex phase where $K=2g\tan\frac{\phi
}{2}\sin\frac{\phi}{2}$ is fulfilled. The right graph shows the chiral current
against $K$ at $\phi=\pi/2$, while the bottom shows the chiral current against
$\phi$ at $K=\sqrt{2}$.$~$In the right one, the chiral current first increases
with $K$ in the vortex phase and then remains unchanged once the critical
value $K_{\text{c}}$ is met, which signifies the Meissner phase. In the bottom
one, the chiral current first rises with $\phi$ in the Meissner phase until a
critical value $\phi_{\text{c}}$ is reached, after which the vortex phase is
entered.}%
\label{fig:chiralcurrentcalculation}%
\end{figure}

The presence of the effective magnetic flux will make the system exhibit the
property of chirality. In detail, the particle currents on both legs differ
from each other. To quantify the difference, we define the chiral particle
current as
\begin{equation}
j_{\text{C}}=j_{\text{L}}-j_{\text{R}}. \label{eq:chiral_current}%
\end{equation}
Here, $j_{d}=(N-1)^{-1}\sum_{l=1}^{N-1}j_{l,l+1}^{\left(  d\right)  } $ with
$d=\mathrm{L,R}$ is the site-averaged current on the particular $d$ leg. In
Fig.~\ref{fig:chiralcurrentcalculation}, the chiral current strength is
plotted as a function of the flux $\phi$ and interleg coupling strength $K$.
The Meissner and vortex phase are separated by a critical boundary, where
$K=2g\tan\frac{\phi}{2}\sin\frac{\phi}{2}$ [see Eq.%
%TCIMACRO{\TeXButton{TeX field}{~}}%
%BeginExpansion
~%
%EndExpansion
(\ref{eq:Kc})] is fulfilled. This boundary corresponds to the degeneracy
transition of the single-particle ground state in the infinite-length case
[see Figs.%
%TCIMACRO{\TeXButton{TeX field}{~}}%
%BeginExpansion
~%
%EndExpansion
\ref{fig:spectrum}(a)-\ref{fig:spectrum}(c)]. For given $K=\sqrt{2}$, the
chiral current first increases as $\phi$ untill reaching its maximum at
$\phi_{\text{c}}=\frac{\pi}{2}$ and then goes down towards zero, while, for
given $\phi=\frac{\pi}{2}$, the chiral current also first increases as $K$
untill reaching its maximum at $K_{\text{c}}=\sqrt{2}$ but never changes
afterwards. The current patterns of the Meissner and vortex phase will be
discussed below.

\subsection{Current patterns in the vortex and Meissner phases}%

%TCIMACRO{\TeXButton{fig:wavefunction6}{\begingroup\begin{figure*}[ptb]
%\includegraphics[height=6.8 cm, clip]{fig7a.eps}
%\includegraphics[height=6.8 cm, clip]{fig7b.eps}
%\caption{(color online).
%Current patterns and photon densities for different values of the interleg
%coupling $K$. Here, the intraleg coupling $g=1$, the flux $\phi=\pi
%/2$ for the left column
%and $-\pi/2$ for the right one, the
%site number $N=20$. The current strength, normalized to the maximum current
%for each $K$, is denoted by the thickness and length of the arrows. The shade
%of the color represents the photon density, which is also normalized to its
%maximum value for each $K$. The flux $\phi=\pm\pi
%/2$ makes the critical value of
%the interleg coupling $K_{\text{c}}=\sqrt{2}%
%$, the value that separates the vortex and
%Meissner phases. In the first row, $K=2.5$, and the currents mainly flow around the
%edges of the ladder, which, forming one large vortex, is called the Meissner
%phase. In the second row, $K=\sqrt{2}%
%$, which is the phase transition point, and the
%current pattern also belongs to Meissner phase. From the third to fourth row where $K=1$
%and $0.5$ successively, the decreasing of $K$ induces the increasing of the vortex number.
%We find that when $\phi$ is flipped from $\pi/2$ to $-\pi/2$,
%the currents also change their directions.
%}
%\label{fig:wavefunction6}
%\end{figure*}
%\endgroup}}%
%BeginExpansion
\begingroup\begin{figure*}[ptb]
\includegraphics[height=6.8 cm, clip]{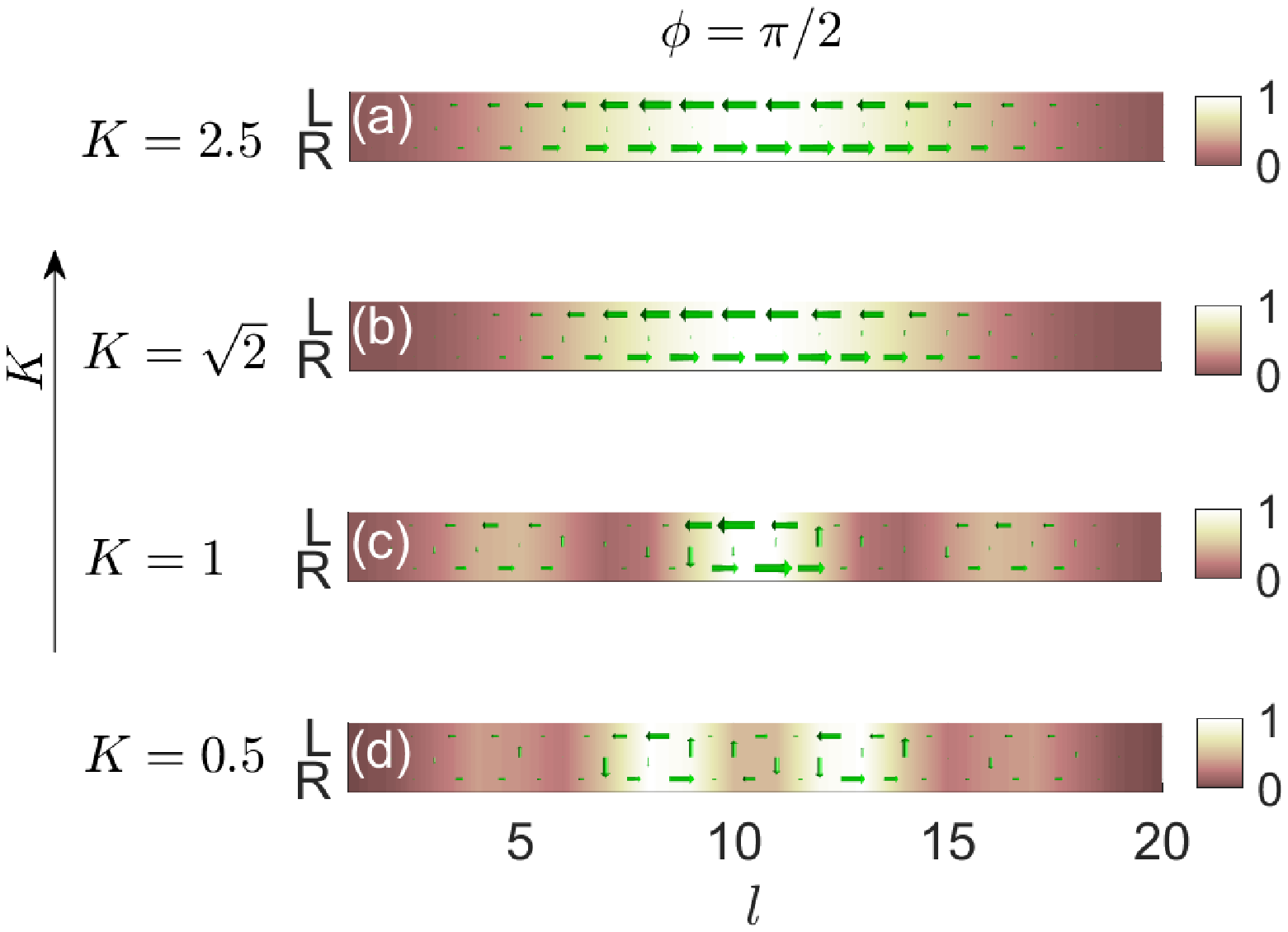}
\includegraphics[height=6.8 cm, clip]{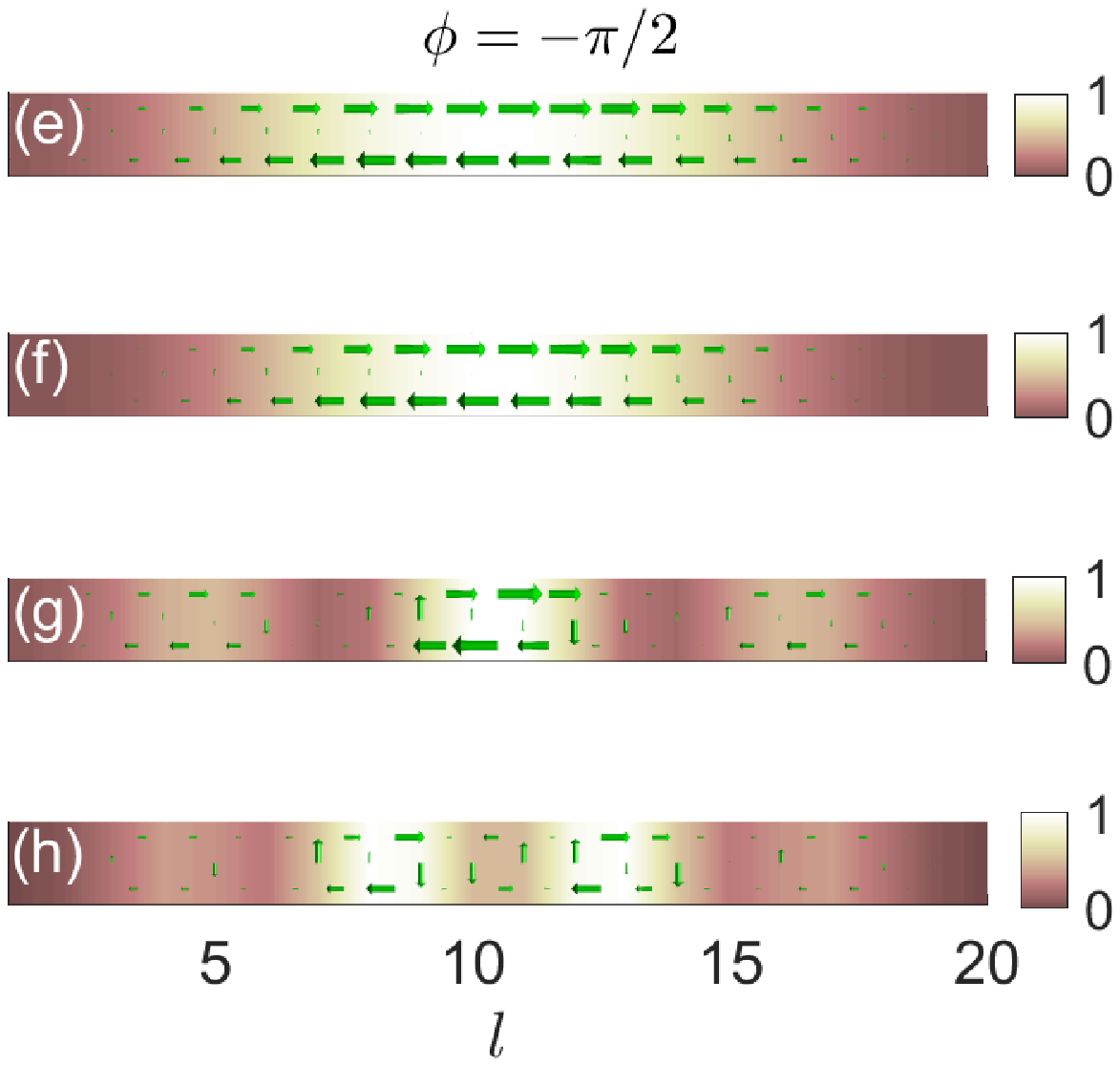}
\caption{(color online).
Current patterns and photon densities for different values of the interleg
coupling $K$. Here, the intraleg coupling $g=1$, the flux $\phi=\pi
/2$ for the left column
and $-\pi/2$ for the right one, the
site number $N=20$. The current strength, normalized to the maximum current
for each $K$, is denoted by the thickness and length of the arrows. The shade
of the color represents the photon density, which is also normalized to its
maximum value for each $K$. The flux $\phi=\pm\pi
/2$ makes the critical value of
the interleg coupling $K_{\text{c}}=\sqrt{2}%
$, the value that separates the vortex and
Meissner phases. In the first row, $K=2.5$, and the currents mainly flow around the
edges of the ladder, which, forming one large vortex, is called the Meissner
phase. In the second row, $K=\sqrt{2}%
$, which is the phase transition point, and the
current pattern also belongs to Meissner phase. From the third to fourth row where $K=1$
and $0.5$ successively, the decreasing of $K$ induces the increasing of the vortex number.
We find that when $\phi$ is flipped from $\pi/2$ to $-\pi/2$,
the currents also change their directions.
}
\label{fig:wavefunction6}
\end{figure*}
\endgroup
%EndExpansion

The difference between vortex and Meissner phases can be intuitively seen from
their individual current patterns in Fig.%
%TCIMACRO{\TeXButton{TeX field}{~}}%
%BeginExpansion
~%
%EndExpansion
\ref{fig:wavefunction6}. In the vortex phase, currents flow around particular
kernels, the number of which is what we define as the vortex number. In the
Meissner phase, the currents only flow along the edges of the ladder, which
can be therefore regarded as a single large vortex. In Fig.%
%TCIMACRO{\TeXButton{TeX field}{~}}%
%BeginExpansion
~%
%EndExpansion
\ref{fig:wavefunction6}, the flux $\phi=\pi/2$ for the left column and
$-\pi/2$ for the right column, the intraleg coupling $g=1$, the site number
$N=20$, and the corresponding critical interleg coupling is $K_{\text{c}%
}=\sqrt{2}$. When $K$ goes down from $2.5$ to the critical value $\sqrt{2}$,
we see no more vortex to occur except the only one circulating around the
edges. However, if $K$ continues to decrease to $1$ and furthermore $0.5$, we
see that more vortices come into being. Moreover, before $K$ reaches $\sqrt
{2}$, the particle density shows no periodical modulation, while, until $K$
reaches $\sqrt{2}$, more modulation periods appear as $K$ is increased. We
mention that due to the effect of the open boundary, the particle density
approaches zero near the chain ends. We also see the change of current
directions when the flux $\phi$ is flipped from $\pi/2$ [see Figs.%
%TCIMACRO{\TeXButton{TeX field}{~}}%
%BeginExpansion
~%
%EndExpansion
\ref{fig:wavefunction6}(a)-\ref{fig:wavefunction6}(d)] to $-\pi/2$ [see Figs.%
%TCIMACRO{\TeXButton{TeX field}{~}}%
%BeginExpansion
~%
%EndExpansion
\ref{fig:wavefunction6}(e)-\ref{fig:wavefunction6}(h)].

\begin{figure}[ptb]
\includegraphics[width=0.4\textwidth, clip]{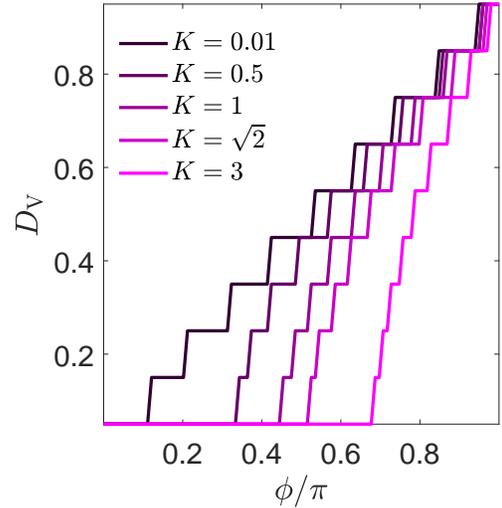}\caption{(color online).
Vortex density $D_{\text{V}}$ as a function of the effective flux $\phi$ and
the interleg coupling $K$ with $N=20$ sites, $g=1$, and the open boundary
condition. For each determined $K$, there is a critical value of the flux
$\phi_{\text{c}}$. Below $\phi_{\text{c}}$, the system is in the Meissner
phase, possessing a constant vortex density $1/N=0.05$, while above
$\phi_{\text{c}}$, the system is in the vortex phase, where the vortex density
increases with the flux $\phi$. }%
\label{fig:vortexdensity4}%
\end{figure}

To numerically quantify the vortex density, i.e., the average vortex number
per lattice site, we now make one count of vortex for a particular plaquette
once such a current pattern as the clockwise or anticlockwise type is present.
Thus, if the total vortex number is $N_{\mathrm{V}}$, vortex density is then
$D_{\mathrm{V}}=N_{\mathrm{V}}/N$. In Fig.%
%TCIMACRO{\TeXButton{TeX field}{~}}%
%BeginExpansion
~%
%EndExpansion
\ref{fig:vortexdensity4}, we have plotted the vortex density $D_{\text{V}}$
against the flux $\phi$ for different values of $K$ with $N=20$, $g=1$, and
the open boundary conditions. For each given $K$, there is a critical value of
the flux $\phi_{\text{c}}$. Below $\phi_{\text{c}}$, the system is in the
Meissner phase, possessing a constant vortex density $1/N=0.05$, while above
$\phi_{\text{c}}$, the system is in the vortex phase, where the vortex density
increases with the flux $\phi$. Since the vortex number must be integers, the
increase of vortex density with $\phi$ is in steps. Besides, the critical flux
$\phi_{\text{c}}$ shifts to the right gradually when $K$ is increased.

\section{Experimental details\label{Sec:ExpDetails}}

\subsection{Generating the single-particle ground state\label{Sec:Gen_SPG}}

To observe the chiral particle current discussed above, we need to generate
the single-particle ground state, i.e., the lowest single-particle energy
state $|\mu_{1}\rangle$. In principle, the cold atoms can be condensed into
one common single-particle state via laser cooling, thus forming the so-called
Bose-Einstein condensate. However, since the number of particles here is not
conserved as that of atoms, the ladder model realized by superconducting qubit
circuits will decay to the ground state (with no particles present) through
sufficient cooling of the conventional dilution refrigerator. Hence, in the
following, we will demonstrate how to generate the single-particle ground
state from the ground state.

\begin{figure}[ptb]
\includegraphics[width=0.38\textwidth, clip]{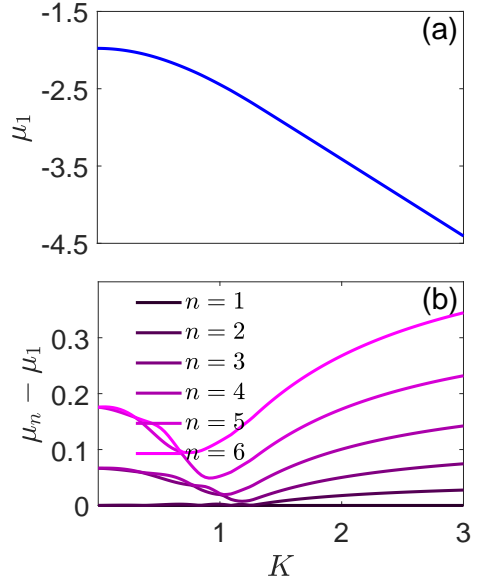}\caption{(color online).
(a) Ground state frequency versus the interleg coupling $K$. (b) Lowest six
eigen frequencies versus $K$ in reference to the ground state frequency. Here,
the flux $\phi=\pi/2$, the site number $N=20$, the intraleg coupling $g=1$,
and the open boundary condition is assumed. We find the critical value of the
interleg coupling $K_{\text{c}}$, below which, the eigen frequencies are
nearly degenerate in pairs. However, above $K_{\text{c}}$, the degeneracy is
broken.}%
\label{fig:E_versus_K}%
\end{figure}

We now discuss a general method that generates the single-particle ground
state from the ground state $\left\vert 0\right\rangle $, and simultaneously
causes no unwanted excitations. In detail, we classically drive the qubits at
all the sites, which appears in Eq.%
%TCIMACRO{\TeXButton{TeX field}{~}}%
%BeginExpansion
~%
%EndExpansion
(\ref{eq:H}) as an additional term
\begin{equation}
\hat{H}_{\text{g}}=\frac{\hbar}{2}\sum_{d=\text{L}}^{\text{R}}\sum_{l=1}%
^{N}\hat{\sigma}_{+}^{\left(  d,l\right)  }B_{d,l}\exp\left(  -i\nu
_{d}t\right)  +\text{H.c.}.
\end{equation}
When we further go to Eq.%
%TCIMACRO{\TeXButton{TeX field}{~}}%
%BeginExpansion
~%
%EndExpansion
(\ref{eq:Hld}), $\hat{H}_{\text{g}}$ is transformed into
\begin{equation}
\hat{H}_{\text{ld,g}}=\frac{\hbar}{2}\sum_{d=\text{L}}^{\text{R}}\sum
_{l=1}^{N}\hat{\sigma}_{+}^{\left(  d,l\right)  }B_{d,l}^{\prime}\exp\left(
-i\epsilon t\right)  +\text{H.c.}. \label{eq:Hldg}%
\end{equation}
Here, the driving strength $B_{d,l}^{\prime}=B_{d,l}J_{0}\left(  \frac{\Omega
}{\delta}\right)  \approx B_{d,l}$, since $\left\vert \Omega/\delta\right\vert
^{2}\ll1$ is satisfied by the parameters in Sec.%
%TCIMACRO{\TeXButton{TeX field}{~}}%
%BeginExpansion
~%
%EndExpansion
\ref{sec:Model}, and the detuning $\epsilon\equiv\nu_{d}-\omega_{d}$ for
$d=\mathrm{L,R}$ can be achieved via carefully tuning $\nu_{d}$. In Fig.%
%TCIMACRO{\TeXButton{TeX field}{~}}%
%BeginExpansion
~%
%EndExpansion
\ref{fig:E_versus_K}, it can be found that the eigenstates are approximately
degenerate in pairs when $K<K_{\text{c}}$, although the approximate degeneracy
is broken when $K>K_{\text{c}}$. Therefore, when we excite the single-particle
ground state $\left\vert \mu_{1}\right\rangle $ from ground state with
$\epsilon=\mu_{1}$, at least the single-particle state $\left\vert \mu
_{2}\right\rangle $ might also be excited and so might the other
single-particle states. 

\begin{figure}[ptb]
\includegraphics[width=0.45\textwidth, clip]{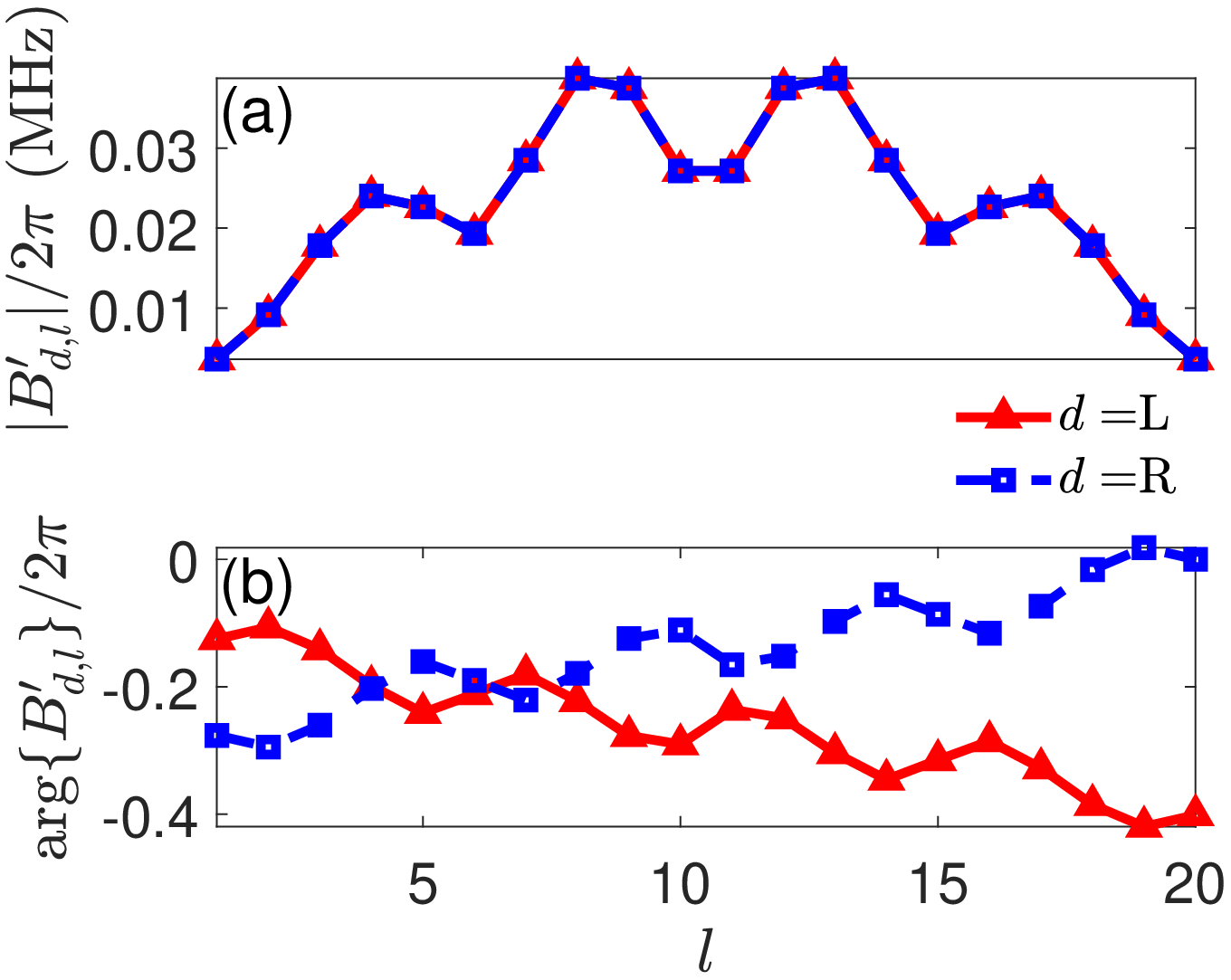}\caption{(color
online). Driving strength $\left\vert B_{d,l}^{\prime}\right\vert /2\pi$ and
phase $\arg\{B_{d,l}^{\prime}\}/2\pi$ at the site $(d,l)$ which is needed to
reach the Rabi frequency $C_{1}/2\pi=1\operatorname{MHz}$ for generating the
single-particle ground state. The solid red (dashed blue) curves marked with
triangles (squares) mean $d=\mathrm{L}$ ($d=\mathrm{R}$). Here, the intraleg
coupling strength $g/2\pi=3.5\operatorname{MHz}$, the interleg coupling
strength $K/2\pi=1.75\operatorname{MHz}$ (such that $K/g=0.5$), the ladder
length $N=20$, and the flux $\phi=\pi/2$ are assumed.}%
\label{fig:DrFld_gen}%
\end{figure}

To overcome this problem, we now make a unitray transformation of the
single-particle creation operator, i.e., $\hat{\sigma}_{+}^{\left(
d,l\right)  }=\sum_{n=1}^{2N}\chi_{d,l}^{\left(  n\right)  \ast}\hat{\Sigma
}_{n}^{+}$, and thus the interaction Hamiltonian in Eq.%
%TCIMACRO{\TeXButton{TeX field}{~}}%
%BeginExpansion
~%
%EndExpansion
(\ref{eq:Hldg}) becomes
\begin{equation}
\hat{H}_{\text{ld,g}}=\frac{\hbar}{2}\sum_{n=1}^{2N}C_{n}\hat{\Sigma}_{n}%
^{+}\exp\left(  -i\epsilon t\right)  +\text{H.c.}. \label{eq:Hld_g_full}%
\end{equation}
Here, the Pauli operator $\hat{\Sigma}_{n}^{+}$ represents the collective
exciations of the qubits, and the driving strength $C_{n}=\sum_{d=\text{L}%
}^{\text{R}}\sum_{l=1}^{N}\chi_{d,l}^{\left(  n\right)  \ast}B_{d,l}^{\prime
}~$can be controlled by the amplitude $B_{d,l}^{\prime}$ (or equivalently,
$B_{d,l}$). To remove the excitations on the single-particle excitation states
(i.e., the states $\left\vert \mu_{n}\right\rangle $ with $n\geq2$), we should
make $C_{n}=0$ for $n\geq2$, which yields the required driving strength
\begin{equation}
B_{d,l}^{\prime}=\sum_{n=1}^{2N}\chi_{d,l}^{\left(  n\right)  }C_{n}%
=\chi_{d,l}^{\left(  1\right)  }C_{1} \label{eq:Bpdl}%
\end{equation}
using the orthonormality condition of $\chi_{d,l}^{\left(  n\right)  }$.
Obviously, the driving fields $B_{d,l}^{\prime}$ must possess the same profile
as the single-particle ground state $\chi_{d,l}^{\left(  1\right)  }$ except
for a scaling factor, i.e., the Rabi frequency $C_{1}$. Then, Eq.%
%TCIMACRO{\TeXButton{TeX field}{~}}%
%BeginExpansion
~%
%EndExpansion
(\ref{eq:Hld_g_full}) can be simplified into
\begin{equation}
\hat{H}_{\text{ld,g}}^{\prime}=\frac{\hbar}{2}C_{1}\exp\left(  -i\epsilon
t\right)  \hat{\Sigma}_{1}^{+}+\mathrm{H.c.},
\end{equation}
where we assume $C_{1}$ is tuned positive. From Eqs.%
%TCIMACRO{\TeXButton{TeX field}{~}}%
%BeginExpansion
~%
%EndExpansion
(\ref{eq:Jordan-Wigner1}) and (\ref{eq:Jordan-Wigner2}), we know that
$\hat{\sigma}_{+}^{\left(  d,l\right)  }\left\vert 0\right\rangle =\hat
{b}_{d,l}^{\dag}\left\vert 0\right\rangle $, thus yielding
\begin{align}
\hat{\Sigma}_{1}^{+}\left\vert 0\right\rangle  &  =\sum_{d=\text{L}}%
^{\text{R}}\sum_{l=1}^{N}\chi_{d,l}^{\left(  1\right)  }\hat{\sigma}%
_{+}^{\left(  d,l\right)  }\left\vert 0\right\rangle \nonumber\\
&  =\sum_{d=\text{L}}^{\text{R}}\sum_{l=1}^{N}\chi_{d,l}^{\left(  1\right)
}\hat{b}_{d,l}^{\dag}\left\vert 0\right\rangle =\left\vert \mu_{1}%
\right\rangle .
\end{align}
Since the single-particle ground state is generated from the ground state, we
then have
\begin{equation}
\hat{H}_{\text{ld,g}}^{\prime}=\frac{\hbar}{2}C_{1}\exp\left(  -i\epsilon
t\right)  \left\vert \mu_{1}\right\rangle
%TCIMACRO{\TeXButton{TeX field}{\!}}%
%BeginExpansion
\!%
%EndExpansion
\left\langle 0\right\vert +\mathrm{H.c.}. \label{eq:Hgld_remove}%
\end{equation}
Thus, the unwanted excitations characterized by $C_{n}$ for $n\geq2$ are all
removed via properly adjusting $B_{d,l}^{\prime}$. If the detuning is further
taken as $\epsilon=\mu_{1}$ as expected, the system will evolve to the state
$\cos\left(  C_{1}t/2\right)
%TCIMACRO{\TeXButton{TeX field}{\!}}%
%BeginExpansion
\!%
%EndExpansion
\left\vert 0\right\rangle -i\sin\left(  C_{1}t/2\right)
%TCIMACRO{\TeXButton{TeX field}{\!}}%
%BeginExpansion
\!%
%EndExpansion
\left\vert \mu_{1}\right\rangle $ in a time duration $t$. Assuming a $\pi$
pulse, i.e., $C_{1}t=\pi$, the single-particle ground state $\left\vert
\mu_{1}\right\rangle $ can be achieved in just one step. If we specify the
intraleg coupling strength $g/2\pi=3.5%
%TCIMACRO{\unit{MHz}}%
%BeginExpansion
\operatorname{MHz}%
%EndExpansion
$, the interleg coupling strength $K/2\pi=1.75%
%TCIMACRO{\unit{MHz}}%
%BeginExpansion
\operatorname{MHz}%
%EndExpansion
$, the ladder length $N=20$, the flux $\phi=\pi/2$, and the detuning
$\epsilon/2\pi=\mu_{1}/2\pi=-210.4%
%TCIMACRO{\unit{MHz}}%
%BeginExpansion
\operatorname{MHz}%
%EndExpansion
$, the driving strength $B_{d,l}^{\prime}$ required to reach the desired Rabi
frequencies $C_{1}/2\pi=1%
%TCIMACRO{\unit{MHz}}%
%BeginExpansion
\operatorname{MHz}%
%EndExpansion
$ and $C_{n}/2\pi=0$ ($n\geq2$) can be shown in Fig.%
%TCIMACRO{\TeXButton{TeX field}{~}}%
%BeginExpansion
~%
%EndExpansion
\ref{fig:DrFld_gen}, which implies a generation time of $0.5%
%TCIMACRO{\unit{\U{3bc}s}}%
%BeginExpansion
\operatorname{\mu s}%
%EndExpansion
$. Besides, we can verify that $\left\vert B_{d,l}^{\prime}\right\vert $ [see
Fig.%
%TCIMACRO{\TeXButton{TeX field}{~}}%
%BeginExpansion
~%
%EndExpansion
\ref{fig:DrFld_gen}(a)] shares the same profile as $\left\vert \chi
_{d,l}^{\left(  1\right)  }\right\vert $ [see Figs.%
%TCIMACRO{\TeXButton{TeX field}{~}}%
%BeginExpansion
~%
%EndExpansion
\ref{fig:wavefunction}(a) and
%TCIMACRO{\TeXButton{TeX field}{~}}%
%BeginExpansion
~%
%EndExpansion
\ref{fig:wavefunction}(b)] except for a scaling factor.

\begin{figure}[ptb]
\includegraphics[width=0.45\textwidth, clip]{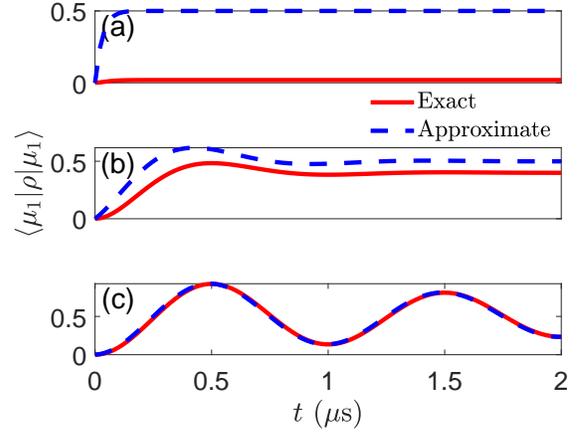}\caption{(color
online). Single-particle ground state fidelity $\left\langle \mu
_{1}\right\vert \!\hat{\rho}\!\left\vert \mu_{1}\right\rangle $ evolving
versus the time $t$ under the effects of environment for the dephasing rate
$\Gamma_{1}$ taking (a) $10C_{1}$, (b) $C_{1}$, and (c) $0.1C_{1}$,
respectively. Here, $C_{1}/2\pi=1\operatorname{MHz}$ is the Rabi frequency.
The relaxation rate takes $\gamma_{1}=0.5\Gamma_{1}$ in all plots. The solid
red (dashed blue) curve denotes the exact solution (the approximate one in the
strong coupling limit $C_{1}\gg\gamma_{1},\Gamma_{1}$). }%
\label{fig:exact-approximate}%
\end{figure}

Having obtained the target Hamiltonian in Eq.%
%TCIMACRO{\TeXButton{TeX field}{~}}%
%BeginExpansion
~%
%EndExpansion
(\ref{eq:Hgld_remove}), we now investigate the effect of the environment on
the state generation process, which is described by the Lindblad master
equation
\begin{equation}
\frac{\text{d}\hat{\rho}}{\text{d}t}=\frac{1}{i\hbar}[\hat{H}_{\text{ld}%
}^{\left(  N\right)  }+\hat{H}_{\text{ld,g}}^{\prime},\hat{\rho}%
]+\mathcal{L}_{\mu1}\left[  \hat{\rho}\right]  . \label{eq:ME_Gen}%
\end{equation}
Here, $\hat{\rho}$ is the density operator of the ladder, $\mathcal{L}_{\mu
1}\left[  \hat{\rho}\right]  $ represents the Lindblad dissipation terms as%
\begin{align}
\mathcal{L}_{\mu1}\left[  \hat{\rho}\right]   &  =-\gamma_{1}\left\vert
\mu_{1}\right\rangle
%TCIMACRO{\TeXButton{TeX field}{\!}}%
%BeginExpansion
\!%
%EndExpansion
\left\langle \mu_{1}\right\vert \left\langle \mu_{1}\right\vert
%TCIMACRO{\TeXButton{TeX field}{\!}}%
%BeginExpansion
\!%
%EndExpansion
\hat{\rho}%
%TCIMACRO{\TeXButton{TeX field}{\!}}%
%BeginExpansion
\!%
%EndExpansion
\left\vert \mu_{1}\right\rangle
%TCIMACRO{\TeXButton{TeX field}{\!}}%
%BeginExpansion
\!%
%EndExpansion
+%
%TCIMACRO{\TeXButton{TeX field}{\!}}%
%BeginExpansion
\!%
%EndExpansion
\gamma_{1}\left\vert 0\right\rangle
%TCIMACRO{\TeXButton{TeX field}{\!}}%
%BeginExpansion
\!%
%EndExpansion
\left\langle 0\right\vert \left\langle 0\right\vert
%TCIMACRO{\TeXButton{TeX field}{\!}}%
%BeginExpansion
\!%
%EndExpansion
\hat{\rho}%
%TCIMACRO{\TeXButton{TeX field}{\!}}%
%BeginExpansion
\!%
%EndExpansion
\left\vert 0\right\rangle
%TCIMACRO{\TeXButton{TeX field}{\!}}%
%BeginExpansion
\!%
%EndExpansion
\nonumber\\
&  -%
%TCIMACRO{\TeXButton{TeX field}{\!}}%
%BeginExpansion
\!%
%EndExpansion
\frac{\Gamma_{1}}{2}\left\vert \mu_{1}\right\rangle
%TCIMACRO{\TeXButton{TeX field}{\!}}%
%BeginExpansion
\!%
%EndExpansion
\left\langle 0\right\vert \left\langle \mu_{1}\right\vert
%TCIMACRO{\TeXButton{TeX field}{\!}}%
%BeginExpansion
\!%
%EndExpansion
\hat{\rho}%
%TCIMACRO{\TeXButton{TeX field}{\!}}%
%BeginExpansion
\!%
%EndExpansion
\left\vert 0\right\rangle
%TCIMACRO{\TeXButton{TeX field}{\!}}%
%BeginExpansion
\!%
%EndExpansion
-%
%TCIMACRO{\TeXButton{TeX field}{\!}}%
%BeginExpansion
\!%
%EndExpansion
\frac{\Gamma_{1}}{2}\left\vert \mu_{1}\right\rangle
%TCIMACRO{\TeXButton{TeX field}{\!}}%
%BeginExpansion
\!%
%EndExpansion
\left\langle 0\right\vert \left\langle \mu_{1}\right\vert
%TCIMACRO{\TeXButton{TeX field}{\!}}%
%BeginExpansion
\!%
%EndExpansion
\hat{\rho}%
%TCIMACRO{\TeXButton{TeX field}{\!}}%
%BeginExpansion
\!%
%EndExpansion
\left\vert 0\right\rangle ,
\end{align}
and $\gamma_{1}$ ($\Gamma_{1}$) is the relaxation (dephasing) rate of the
single-particle ground state $\left\vert \mu_{1}\right\rangle $. Using Eq.%
%TCIMACRO{\TeXButton{TeX field}{~}}%
%BeginExpansion
~%
%EndExpansion
(\ref{eq:ME_Gen}), we can find the exact solution of $\left\langle \mu
_{1}\right\vert
%TCIMACRO{\TeXButton{TeX field}{\!}}%
%BeginExpansion
\!%
%EndExpansion
\hat{\rho}\left(  t\right)
%TCIMACRO{\TeXButton{TeX field}{\!}}%
%BeginExpansion
\!%
%EndExpansion
\left\vert \mu_{1}\right\rangle $ (see Appendix.%
%TCIMACRO{\TeXButton{TeX field}{~}}%
%BeginExpansion
~%
%EndExpansion
\ref{Append:exact}), i.e., the fidelity of the single-particle ground state at
the time $t$. However, in the strong coupling limit ($C_{1}\gg\gamma_{1}$,
$\Gamma_{1}$), the generation fidelity can be approximated as
\begin{equation}
\left\langle \mu_{1}\right\vert
%TCIMACRO{\TeXButton{TeX field}{\!}}%
%BeginExpansion
\!%
%EndExpansion
\hat{\rho}%
%TCIMACRO{\TeXButton{TeX field}{\!}}%
%BeginExpansion
\!%
%EndExpansion
\left\vert \mu_{1}\right\rangle =\frac{1}{2}\left[  1-e^{-\frac{1}{2}\left(
\gamma_{1}+\frac{\Gamma_{1}}{2}\right)  t}\cos\left(  C_{1}t\right)  \right]
.
\end{equation}
Suppose the relaxation (dephasing) rate of the qubit at the site $\left(
d,l\right)  $ is $\gamma_{d,l}$ ($\Gamma_{d,l}$), then $\gamma_{1}$ and
$\Gamma_{1}$ can be estimated by
\begin{equation}
\gamma_{1}=\sum_{d,l}|\chi_{d,l}^{\left(  1\right)  }|^{2}\gamma_{d,l}\text{
and }\Gamma_{1}=\sum_{d,l}|\chi_{d,l}^{\left(  1\right)  }|^{2}\Gamma_{d,l}.
\end{equation}
We consider homogeneous qubit decay rates, e.g., $\gamma_{d,l}/2\pi\equiv0.05%
%TCIMACRO{\unit{MHz}}%
%BeginExpansion
\operatorname{MHz}%
%EndExpansion
$ and $\Gamma_{d,l}/2\pi\equiv0.1%
%TCIMACRO{\unit{MHz}}%
%BeginExpansion
\operatorname{MHz}%
%EndExpansion
$, while other parameters remain unchanged. Then, after a $\pi~$pulse, the
fidelity is about $\left\langle \mu_{1}\right\vert
%TCIMACRO{\TeXButton{TeX field}{\!}}%
%BeginExpansion
\!%
%EndExpansion
\hat{\rho}(\frac{\pi}{C_{1}})%
%TCIMACRO{\TeXButton{TeX field}{\!}}%
%BeginExpansion
\!%
%EndExpansion
\left\vert \mu_{1}\right\rangle =$ $0.9273$. In
Fig.~\ref{fig:exact-approximate}, we have shown the exact solution and the
approximate one for the weak ($\Gamma_{1}=10C_{1}$), critical ($\Gamma
_{1}=C_{1}$), and strong ($\Gamma_{1}=0.1C$) coupling, where good agreement is
found in the last case.

\subsection{Measurement scheme\label{sec:Measurement Scheme}}

To observe the vortex-Meissner phase transition, one indispensable issue is to
measure the particle currents between a pair of adjacent sites. In
superconducting quantum circuits, the qubit state can be dispersively read out
by a microwave resonator, which enables us to extract the particle current
from the Rabi oscillation between the pair of adjacent sites. To achieve this,
we can tune the energy levels of the flux qubits that connect to the pair of
sites we concentrate on such that both sites are decoupled from the others.
For example, to investigate the Rabi oscillation between $\left(
\text{L},l\right)  $ and $\left(  \text{L},l+1\right)  $, we can tune the flux
qubits at the sites $\left(  \text{L},l-1\right)  $, $\left(  \text{L}%
,l+2\right)  $, $\left(  \text{R},l\right)  $, and $\left(  \text{R}%
,l+1\right)  $ such that they are decoupled from the ones at $\left(
\text{L},l\right)  $ and $\left(  \text{L},l+1\right)  $. Then, the bare
Hamiltonian that governs the evolution of the adjacent sites $\left(
\text{L},l\right)  $ and $\left(  \text{L},l+1\right)  $ can be given by
$\hat{H}_{\text{L},l}=-\hbar g\hat{\sigma}_{+}^{\left(  \text{L},l+1\right)
}\hat{\sigma}_{-}^{\left(  \text{L},l\right)  }+\mathrm{H.c..}$ Differently
from the cold atoms in optical lattices, the particles stored in the flux
qubits suffer the relaxation rates $\gamma_{d,l}$ and dephasing rates
$\Gamma_{d,l}$ for the site $\left(  d,l\right)  $. Thus, the interaction
between the qubits at $\left(  \text{L},l\right)  $ and $\left(
\text{L},l+1\right)  $ should also be described by the Lindblad master
equation, i.e.,
\begin{equation}
\frac{\text{d}\hat{\rho}_{\text{L},l}}{\text{d}t}=\frac{[\hat{H}_{\text{L}%
,l},\hat{\rho}_{\text{L},l}]}{i\hbar}+\mathcal{L}_{\text{L,}l}\left[
\hat{\rho}_{\text{L},l}\right]  +\mathcal{L}_{\text{L,}l+1}\left[  \hat{\rho
}_{\text{L},l}\right]  .
\end{equation}
Here, $\hat{\rho}_{\text{L},l}=\sum_{l_{1}=l}^{l+1}\sum_{l_{2}=l}%
^{l+1}\left\vert \text{L,}l_{1}\right\rangle
%TCIMACRO{\TeXButton{TeX field}{\!}}%
%BeginExpansion
\!%
%EndExpansion
\left\langle \text{L,}l_{1}\right\vert \hat{\rho}\left\vert \text{L,}%
l_{2}\right\rangle
%TCIMACRO{\TeXButton{TeX field}{\!}}%
%BeginExpansion
\!%
%EndExpansion
\left\langle \text{L,}l_{2}\right\vert $ is the subspace truncation of the
global density operator $\hat{\rho}$, and the Lindblad terms
\begin{align}
\mathcal{L}_{\mathrm{L},l}%
%TCIMACRO{\TeXButton{TeX field}{\!}}%
%BeginExpansion
\!%
%EndExpansion
\left[  \hat{\rho}_{\text{L},l}\right]   &  =-\gamma_{\mathrm{L},l}\left\vert
\mathrm{L},l\right\rangle
%TCIMACRO{\TeXButton{TeX field}{\!}}%
%BeginExpansion
\!%
%EndExpansion
\left\langle \mathrm{L},l\right\vert
%TCIMACRO{\TeXButton{TeX field}{\!}}%
%BeginExpansion
\!%
%EndExpansion
\left\langle \mathrm{L},l\right\vert
%TCIMACRO{\TeXButton{TeX field}{\!}}%
%BeginExpansion
\!%
%EndExpansion
\hat{\rho}_{\text{L},l}%
%TCIMACRO{\TeXButton{TeX field}{\!}}%
%BeginExpansion
\!%
%EndExpansion
\left\vert \mathrm{L},l\right\rangle \nonumber\\
&  +\gamma_{\mathrm{L},l}\left\vert 0\right\rangle
%TCIMACRO{\TeXButton{TeX field}{\!}}%
%BeginExpansion
\!%
%EndExpansion
\left\langle 0\right\vert
%TCIMACRO{\TeXButton{TeX field}{\!}}%
%BeginExpansion
\!%
%EndExpansion
\left\langle 0\right\vert
%TCIMACRO{\TeXButton{TeX field}{\!}}%
%BeginExpansion
\!%
%EndExpansion
\hat{\rho}_{\text{L},l}%
%TCIMACRO{\TeXButton{TeX field}{\!}}%
%BeginExpansion
\!%
%EndExpansion
\left\vert 0\right\rangle -\frac{\Gamma_{\mathrm{L},l}}{2}\left\vert
0\right\rangle
%TCIMACRO{\TeXButton{TeX field}{\!}}%
%BeginExpansion
\!%
%EndExpansion
\left\langle \mathrm{L},l\right\vert
%TCIMACRO{\TeXButton{TeX field}{\!}}%
%BeginExpansion
\!%
%EndExpansion
\left\langle 0\right\vert
%TCIMACRO{\TeXButton{TeX field}{\!}}%
%BeginExpansion
\!%
%EndExpansion
\hat{\rho}_{\text{L},l}%
%TCIMACRO{\TeXButton{TeX field}{\!}}%
%BeginExpansion
\!%
%EndExpansion
\left\vert \mathrm{L},l\right\rangle \nonumber\\
&  -\frac{\Gamma_{\mathrm{L},l}}{2}\left\vert \mathrm{L},l\right\rangle
%TCIMACRO{\TeXButton{TeX field}{\!}}%
%BeginExpansion
\!%
%EndExpansion
\left\langle 0\right\vert
%TCIMACRO{\TeXButton{TeX field}{\!}}%
%BeginExpansion
\!%
%EndExpansion
\left\langle \mathrm{L},l\right\vert \hat{\rho}_{\text{L},l}%
%TCIMACRO{\TeXButton{TeX field}{\!}}%
%BeginExpansion
\!%
%EndExpansion
\left\vert 0\right\rangle
\end{align}
represent the dissipation into the environment. In the limit of strong
coupling (i.e., $g\gg\gamma_{\mathrm{L},l},\Gamma_{\mathrm{L},l}$), the
population difference between $\left(  \text{L},l+1\right)  $ and $\left(
\text{L},l\right)  $, which defined by $P_{\text{L,}l}\left(  t\right)
=\left\langle \text{L},l+1\right\vert
%TCIMACRO{\TeXButton{TeX field}{\!}}%
%BeginExpansion
\!%
%EndExpansion
\hat{\rho}_{\text{L},l}%
%TCIMACRO{\TeXButton{TeX field}{\!}}%
%BeginExpansion
\!%
%EndExpansion
\left\vert \text{L},l+1\right\rangle -\left\langle \text{L},l\right\vert
%TCIMACRO{\TeXButton{TeX field}{\!}}%
%BeginExpansion
\!%
%EndExpansion
\hat{\rho}_{\text{L},l}%
%TCIMACRO{\TeXButton{TeX field}{\!}}%
%BeginExpansion
\!%
%EndExpansion
\left\vert \text{L},l\right\rangle $, can be obtained using the Lindblad
master equation as
\begin{equation}
P_{\text{L,}l}\left(  t\right)
%TCIMACRO{\TeXButton{TeX field}{\!}}%
%BeginExpansion
\!%
%EndExpansion
=%
%TCIMACRO{\TeXButton{TeX field}{\!}}%
%BeginExpansion
\!%
%EndExpansion
e^{-\tilde{\gamma}_{\text{L,}l}t}%
%TCIMACRO{\TeXButton{TeX field}{\!}}%
%BeginExpansion
\!%
%EndExpansion
\left[  \cos\left(  \tilde{g}t\right)
%TCIMACRO{\TeXButton{TeX field}{\!}}%
%BeginExpansion
\!%
%EndExpansion
P_{\text{L,}l}%
%TCIMACRO{\TeXButton{TeX field}{\!}}%
%BeginExpansion
\!%
%EndExpansion
\left(  0\right)  +\sin\left(  \tilde{g}t\right)
%TCIMACRO{\TeXButton{TeX field}{\!}}%
%BeginExpansion
\!%
%EndExpansion
\frac{j_{l,l+1}^{\left(  \text{L}\right)  }}{g}\right]  ,
\label{eq:PopulationDifference}%
\end{equation}
where $\tilde{\gamma}_{\text{L},l}=\left(  \gamma_{\text{L,}l}+\gamma
_{\text{L,}l+1}+\Gamma_{\text{L,}l}+\Gamma_{\text{L,}l+1}\right)  /4$ and
$\tilde{g}=2g$. Now, we can confidently assert that the particle current
$j_{l,l+1}^{\left(  \text{L}\right)  }$ can be extracted from the population
difference after fitting the measured data using Eq.%
%TCIMACRO{\TeXButton{TeX field}{~}}%
%BeginExpansion
~%
%EndExpansion
(\ref{eq:PopulationDifference}). The discussions made above can also apply to
extracting the particle current on the R leg, for which, the population
difference between $\left(  \text{R},l\right)  $ and $\left(  \text{R}%
,l+1\right)  $ is namely Eq.%
%TCIMACRO{\TeXButton{TeX field}{~}}%
%BeginExpansion
~%
%EndExpansion
(\ref{eq:PopulationDifference}) with the subscript L replaced with R.
Similarly, the population difference between $\left(  \text{R},l\right)  $ and
$\left(  \text{L},l\right)  $ is
\begin{equation}
P_{\text{LR,}l}%
%TCIMACRO{\TeXButton{TeX field}{\!}}%
%BeginExpansion
\!%
%EndExpansion
\left(  t\right)  =e^{-\tilde{\gamma}_{\text{LR,}l}t}%
%TCIMACRO{\TeXButton{TeX field}{\!\!}}%
%BeginExpansion
\!\!%
%EndExpansion
\left[  \cos(\tilde{K}t)P_{\text{LR,}l}%
%TCIMACRO{\TeXButton{TeX field}{\!}}%
%BeginExpansion
\!%
%EndExpansion
\left(  0\right)  +\sin(\tilde{K}t)\frac{j_{\text{LR},l}}{K}\right]
%TCIMACRO{\TeXButton{TeX field}{\!}}%
%BeginExpansion
\!%
%EndExpansion
\text{,}%
\end{equation}
where $\tilde{\gamma}_{\text{LR},l}=\left(  \gamma_{\text{L,}l}+\gamma
_{\text{R,}l}+\Gamma_{\text{L,}l}+\Gamma_{\text{R,}l}\right)  /4$, $\tilde
{K}=2K$, and strong interleg coupling (i.e., $K\gg\gamma_{\mathrm{L},l}%
,\Gamma_{\mathrm{L},l}$) has been assumed.

\begin{figure}[ptb]
\includegraphics[width=0.49\textwidth, clip]{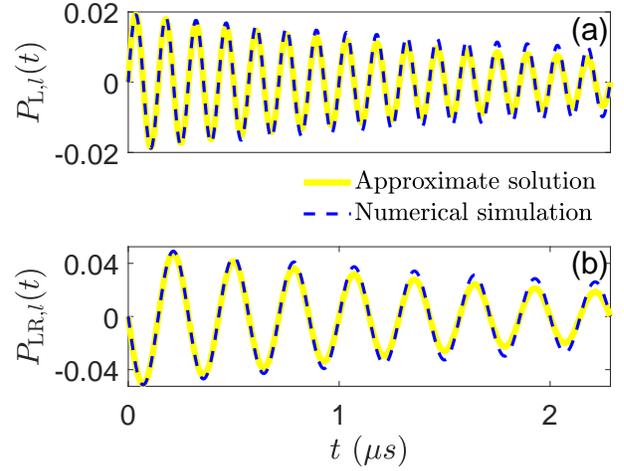}\caption{(color
online). Population difference (a) $P_{\text{L,}l}\left(  t\right)  $ between
the site ($\mathrm{L},l+1$) and ($\mathrm{L},l$), and (b) $P_{\text{LR}%
,l}\left(  t\right)  $ between the site ($\mathrm{R},l$) and ($\mathrm{L},l$)
evolving against the time $t$. The solid yellow (dashed blue) curves
represents the exact numerical simulation results (approximate solutions) in
the strong coupling limit. Here, we specify the chain length $N=20$, the
lattice index $l=N/2=10$, the intraleg coupling strength $g/2\pi
=3.5\operatorname{MHz}$, the interleg coupling strength $K/2\pi
=1.75\operatorname{MHz}$, and the decay rates at the site $( d,l^{\prime}) $
$\gamma_{d,l^{\prime}}/2\pi\equiv0.05\operatorname{MHz}$ and $\Gamma
_{d,l^{\prime}}/2\pi\equiv0.1\operatorname{MHz}$. The corresponding particle
current is (a) $j_{l,l+1}^{\left(  \text{L}\right)  }=0.43\operatorname{MHz}$
and (b) $j_{\text{LR},l}=-0.5785\operatorname{MHz}$.}%
\label{fig:Measurement}%
\end{figure}

In Fig.%
%TCIMACRO{\TeXButton{TeX field}{~}}%
%BeginExpansion
~%
%EndExpansion
\ref{fig:Measurement}, we have intuitively presented the population difference
$P_{\text{L,}l}\left(  t\right)  $ and $P_{\text{LR,}l}\left(  t\right)  $
evolving as the time for $l$ taking $N/2$, with the chain length $N=20$, the
intraleg coupling strength $g/2\pi=3.5%
%TCIMACRO{\unit{MHz}}%
%BeginExpansion
\operatorname{MHz}%
%EndExpansion
$, the interleg coupling strength $K/2\pi=1.75%
%TCIMACRO{\unit{MHz}}%
%BeginExpansion
\operatorname{MHz}%
%EndExpansion
$ (such that $K/g=0.5$), the effective magnetic flux $\phi=\pi/2$, and the
decay rates $\gamma_{d,l^{\prime}}/2\pi\equiv0.05%
%TCIMACRO{\unit{MHz}}%
%BeginExpansion
\operatorname{MHz}%
%EndExpansion
$ and $\Gamma_{d,l^{\prime}}/2\pi\equiv0.1%
%TCIMACRO{\unit{MHz}}%
%BeginExpansion
\operatorname{MHz}%
%EndExpansion
$. The corresponding particle current is $j_{l,l+1}^{\left(  \text{L}\right)
}=0.43%
%TCIMACRO{\unit{MHz}}%
%BeginExpansion
\operatorname{MHz}%
%EndExpansion
$ and $j_{\text{LR},l}=-0.5785%
%TCIMACRO{\unit{MHz}}%
%BeginExpansion
\operatorname{MHz}%
%EndExpansion
$. We find that, in the strong coupling limit, the approximate analytical
solutions (solid blue) agree very well with the exact numerical simulation
results (dashed green), especially in the first few periods. However, when
time goes longer, some deviation is exhibited from the approximate and
numerical results. Thus, to improve accuracy of measurement, we advice to fit
the data from the first few oscillation periods.

Having measured the particle currents between adjacent sites, we can then
calculate the chiral current given in Eq.%
%TCIMACRO{\TeXButton{TeX field}{~}}%
%BeginExpansion
~%
%EndExpansion
(\ref{eq:chiral_current}), which enables us to obtain the vortex-Meissner
phase transition diagram for different interleg coupling strength $K$ and
effective magnetic flux $\phi$ (see Fig.%
%TCIMACRO{\TeXButton{TeX field}{~}}%
%BeginExpansion
~%
%EndExpansion
\ref{fig:chiralcurrentcalculation}). The current patterns (see Fig.%
%TCIMACRO{\TeXButton{TeX field}{~}}%
%BeginExpansion
~%
%EndExpansion
\ref{fig:wavefunction6}) can also be obtained from the particle currents,
which enables us to calculate the vortex density for different $K$ and $\phi$
(see Fig.%
%TCIMACRO{\TeXButton{TeX field}{~}}%
%BeginExpansion
~%
%EndExpansion
\ref{fig:vortexdensity4}). In a word, the vortex-Meissner phase transition can
be determined from the measured data of particle currents between adjacent sites.

\section{Conclusion\label{sec:DC}}

We have introduced a circuit scheme on how to construct the two-leg fermionic
ladder with X-shape gradiometer superconducting flux qubits. In such a scheme,
we have shown that with two-tone driving fields, an artificial effective
magnetic flux can be generated for each plaquette, which can be felt by the
\textquotedblleft fermionic\textquotedblright\ particle and thus affects its
motion. Compared with the previous method for generating effective magnetic
flux without the aid of couplers%
%TCIMACRO{\TeXButton{TeX field}{~}}%
%BeginExpansion
~%
%EndExpansion
\cite{Alaeian2019PRA}, our method does not require the qubit circuit poessess
a weak anharmonicity but on the contrary has a simple analytical expression in
the strong anharmonicity regime. The maintenance of anharmonicity (or
nonlinearity) is crucial, since it is indispensable for demonstrating quantum
behaviors
%TCIMACRO{\TeXButton{TeX field}{~}}%
%BeginExpansion
~%
%EndExpansion
\cite{Wendin2007LTP}.

Via modifying the interleg coupling strength or the effective magnetic flux,
both of which are tunable via adjusting the phases of the classical driving
fields, the vortex-Meissner phase transition can in principle be observed in
the single-particle ground state, which originates from the competition
between the two parameters. In the vortex phase, the number of vortex kernels
are more than one, while in the Meissner phase, there is only one large
vortex, with the currents mainly flowing around the boundaries of the ladder.
The phase transition boundary is analytically given. Besides, the wave
functions, current patterns, and quasimomentum distributions in both phases
are exhaustively discussed. The vortex densities for different parameters have
also been presented.

Since the vortex and Meissner phases are discussed in the single-particle
ground state, which is not the (global) ground state, we have proposed a
method on how to generate the single-particle ground state from the ground
state with just a one-step $\pi$ pulse realized by simultaneously driving all
the qubits and meanwhile cause no undesired excitations. The requied driving
fields should share the same profile as the wave function of the
single-particle ground state except for a scaling factor, the Rabi frequency
of generation.

We have shown that the particle currents between the two adjacent sites can be
extracted from the Rabi oscillations between them, assuming the other sites
connected to them are tuned to decouple. The detailed analytical expression
has been given for fitting the experimentally measured data. The
particle-current measurement between adjacent sites enables the calculation of
chiral particle currents, which is critical for experimentally determining the
vortex-Meissner phase transition.

For strictness, the effects of the environment are also considered for
generating the single-particle ground state and measuring the particle
currents between the adjacent sites. To guarantee the generation fidelity and
measurement accuracy, we find that the sample needs to reach the strong
coupling regime, i.e., the coupling strength should be much larger than the
decay rates. This condition, we think, should not be very difficult to met,
since the ultrastrong coupling%
%TCIMACRO{\TeXButton{TeX field}{~\cite
%{Niemczyk2010NP,Forn2016NP,Yoshihara2017NP}} }%
%BeginExpansion
~\cite{Niemczyk2010NP,Forn2016NP,Yoshihara2017NP}
%EndExpansion
and decoherence time about tens of microseconds%
%TCIMACRO{\TeXButton{TeX field}{~\cite{Yan2016NC,Abdurakhimov2019APL}} }%
%BeginExpansion
~\cite{Yan2016NC,Abdurakhimov2019APL}
%EndExpansion
have both been reported in flux qubit systems.

\section{Acknowledgments}

We are grateful to Wei Han for helpful discussions. Y. J. Z. is supported by
National Natural Science Foundation of China (NSFC) under grants No.s 11904013
and 11847165. Y.X.L. is supported by the Key-Area Research and Development
Program of GuangDong Province under Grant No.2018B030326001, the National
Basic Research Program(973) of China under Grant No. 2017YFA0304304. W. M. L.
was supported by the National Key R\&D Program of China under grants No.
2016YFA0301500, NSFC under grants Nos. 61835013, Strategic Priority Research
Program of the Chinese Academy of Sciences under grants Nos. XDB01020300, XDB21030300.

\appendix

\section{Periodical modulation of the qubit frequency\label{appendix:PMQF}}

Now, we investigate the periodical modulation of a qubit frequency with a
general qubit (e.g., flux qubit, transmon qubit, etc) with multiple energy
levels. The qubit Hamiltonian with two-tone driving fields can be represented as%

\begin{equation}
\hat{H}_{\text{q}}=\hat{H}_{0}+\frac{\hbar}{2}\sum_{n}^{N-1}\sum_{j=1}%
^{2}\left(  \hat{\sigma}_{n+1,n}\Omega_{jn}e^{-i\tilde{\omega}_{j}%
t}+\mathrm{H.c.}\right)  , \label{eq:Hq_append}%
\end{equation}
where $\hat{H}_{0}=\sum_{n}\hbar\omega_{qn}\hat{\sigma}_{nn}$ and $\hat
{\sigma}_{nn}=\left\vert n\right\rangle \left\langle n\right\vert $
($\hat{\sigma}_{n+1,n}=\left\vert n+1\right\rangle \left\langle n\right\vert
)$ is the projection (ladder) operator. In the interaction picture defined by
$\hat{U}_{0}\left(  t\right)  =e^{-i\hat{H}_{0}t}$, the Hamiltonian $\hat
{H}_{\text{q}}$ is transformed into
\begin{equation}
\hat{H}_{\text{I}}\left(  t\right)  =\frac{\hbar}{2}\sum_{n}^{N-1}\sum
_{j=1}^{2}\left(  \hat{\sigma}_{n+1,n}\Omega_{jn}e^{-i\delta_{jn}%
t}+\mathrm{H.c.}\right)  \text{,}%
\end{equation}
where $\delta_{jn}=\tilde{\omega}_{j}-\left(  \omega_{q,n+1}-\omega
_{q,n}\right)  $ is the detuning between the driving field and the applied
energy level.

To derive the effective Hamiltonian, we employ the second-order perturbation
theory in the large-detuning regime $\left\vert \Omega_{jn}/\delta_{j^{\prime
}n}\right\vert ^{2}\ll1$, thus resulting in the evolution operator in the
interaction as
\begin{align}
\hat{U}_{\text{I}}\left(  t\right)  \cong &  1+\frac{1}{i\hbar}\int_{0}%
^{t}\text{d}t^{\prime}\hat{H}_{\text{I}}\left(  t^{\prime}\right) \nonumber\\
&  +\frac{1}{\left(  i\hbar\right)  ^{2}}\int_{0}^{t}\text{d}t^{\prime}\hat
{H}_{\text{I}}\left(  t^{\prime}\right)  \int_{0}^{t^{\prime}}\hat
{H}_{\text{I}}\left(  t^{\prime\prime}\right)  \text{d}t^{\prime\prime}.
\label{eq:pert}%
\end{align}
In the time scale $t\gtrsim\frac{1}{\left\vert \Omega_{jn}\right\vert }$,
which satisfies $t\gg\frac{1}{\left\vert \delta_{jn}\right\vert }$, the
fast-oscillating term (i.e., the first-order perturbative term) in Eq.%
%TCIMACRO{\TeXButton{TeX field}{~}}%
%BeginExpansion
~%
%EndExpansion
(\ref{eq:pert}) can be neglected, thus resulting in%
\begin{align}
\hat{U}_{\text{I}}\cong &  1%
%TCIMACRO{\TeXButton{TeX field}{\!}}%
%BeginExpansion
\!%
%EndExpansion
+\frac{1}{i^{2}}\sum_{n=0}^{N-1}\int_{0}^{t}\text{d}t^{\prime}\sum_{j=1}%
^{2}\frac{\left\vert \Omega_{jn}\right\vert ^{2}}{4}\left(  \frac{\hat{\sigma
}_{n+1,n+1}}{i\delta_{jn}}-\frac{\hat{\sigma}_{n,n}}{i\delta_{jn}}\right)
\nonumber\\
&  +\frac{1}{4i^{2}}\sum_{n=0}^{N-1}\int_{0}^{t}\text{d}t^{\prime}\left(
\frac{O_{n}}{i\delta_{1n}}\hat{\sigma}_{n+1,n+1}-\frac{O_{n}^{\ast}}%
{i\delta_{1n}}\hat{\sigma}_{n,n}\right) \nonumber\\
&  +%
%TCIMACRO{\TeXButton{TeX field}{\!}}%
%BeginExpansion
\!%
%EndExpansion
\frac{1}{4i^{2}}%
%TCIMACRO{\TeXButton{TeX field}{\!}}%
%BeginExpansion
\!%
%EndExpansion
\sum_{n=0}^{N-1}%
%TCIMACRO{\TeXButton{TeX field}{\!}}%
%BeginExpansion
\!%
%EndExpansion
\int_{0}^{t}\text{d}t^{\prime}%
%TCIMACRO{\TeXButton{TeX field}{\!}}%
%BeginExpansion
\!%
%EndExpansion
\left(  \frac{O_{n}^{\ast}}{i\delta_{2n}}\hat{\sigma}_{n+1,n+1}-\hat{\sigma
}_{n,n}\frac{O_{n}}{i\delta_{2n}}\right)
%TCIMACRO{\TeXButton{TeX field}{\!}}%
%BeginExpansion
\!%
%EndExpansion
,
\end{align}
where the symbol $O_{n}\equiv O_{n}\left(  t\right)  =$ $\Omega_{1n}^{\ast
}\Omega_{2n}e^{-i\tilde{\delta}t}$ and the detuning $\tilde{\delta}=$
$\delta_{2n}-\delta_{1n}=\tilde{\omega}_{2}-\tilde{\omega}_{1}$. Assuming
$|\tilde{\delta}|\ll\left\vert \delta_{jn}\right\vert $, which implies
$\delta_{1n}\approx\delta_{2n}$, we can obtain the effective Hamiltonian using
the relation $H_{\text{I,eff}}=i\hbar\partial_{t}U_{\text{I}}\left(  t\right)
$ as%
\begin{align}
\hat{H}_{\text{I,eff}}  &  =\sum_{j=1}^{2}\frac{\hbar\left\vert \Omega
_{j0}\right\vert ^{2}}{4\delta_{j0}}\hat{\sigma}_{00}\nonumber\\
&  +\sum_{n=0}^{N-1}\sum_{j}\left(  \frac{\hbar\left\vert \Omega
_{j,n+1}\right\vert ^{2}}{4\delta_{j,n+1}}-\frac{\hbar\left\vert \Omega
_{jn}\right\vert ^{2}}{4\delta_{jn}}\right)  \hat{\sigma}_{n+1,n+1}\nonumber\\
&  -\sum_{n=0}^{N-1}\frac{\hbar}{2}\frac{\left\vert \Omega_{1n}\Omega
_{2n}\right\vert }{\delta_{1n}}\hat{\sigma}_{n+1,n+1}\cos\left(  \tilde
{\delta}t+\phi_{n}\right) \nonumber\\
&  +\sum_{n=0}^{N-1}\frac{\hbar}{2}\frac{\left\vert \Omega_{1n}\Omega
_{2n}\right\vert }{\delta_{1n}}\hat{\sigma}_{nn}\cos\left(  \tilde{\delta
}t+\phi_{n}\right)  ,
\end{align}
where we have defined $\phi_{1n}-\phi_{2n}\equiv\phi_{n}$. Omitting an
irrelevant constant, the effective Hamiltonian can be further represented as%
\begin{equation}
\hat{H}_{\text{I,eff}}\cong\sum_{n=1}^{N}\hbar\left[  \nu_{n}+\eta_{n}%
\cos\left(  \tilde{\delta}t+\phi_{n-1}\right)  \right]  \hat{\sigma}_{n,n},
\end{equation}
where $\nu_{n}$ is the Stark shift$\,$and $\eta_{n}$ is the periodical
modulation strength:
\begin{align}
\nu_{n}  &  =\sum_{j=1}^{2}\frac{\left\vert \Omega_{jn}\right\vert ^{2}%
}{4\delta_{jn}}-\frac{\left\vert \Omega_{j,n-1}\right\vert ^{2}}%
{4\delta_{j,n-1}}-\frac{\left\vert \Omega_{j0}\right\vert ^{2}}{4\delta_{j0}%
},\\
\eta_{n}  &  =%
%TCIMACRO{\TeXButton{TeX field}{\!}}%
%BeginExpansion
\!%
%EndExpansion
\frac{1}{2}%
%TCIMACRO{\TeXButton{TeX field}{\!}}%
%BeginExpansion
\!%
%EndExpansion
\left(  \frac{\left\vert \Omega_{1n}\Omega_{2n}\right\vert }{\delta_{1n}}%
%TCIMACRO{\TeXButton{TeX field}{\!}}%
%BeginExpansion
\!%
%EndExpansion
-\frac{\left\vert \Omega_{1,n-1}\Omega_{2,n-1}\right\vert }{\delta_{1,n-1}}%
%TCIMACRO{\TeXButton{TeX field}{\!}}%
%BeginExpansion
\!%
%EndExpansion
-\frac{\left\vert \Omega_{10}\Omega_{20}\right\vert }{\delta_{10}}\right)
%TCIMACRO{\TeXButton{TeX field}{\!}}%
%BeginExpansion
\!%
%EndExpansion
.
\end{align}

Returning to the original frame, the effective Hamiltonian is transformed into
the form%
\begin{equation}
\hat{H}_{\text{eff}}\cong\sum_{n=1}^{N}\hbar\left[  \tilde{\omega}_{qn}%
+\eta_{n}\cos\left(  \tilde{\delta}t+\phi_{n-1}\right)  \right]  \hat{\sigma
}_{n,n},
\end{equation}
where $\tilde{\omega}_{qn}=\omega_{qn}+\nu_{n}$. In the large-detuning regime,
the Stark shift $\nu_{n}$ is a small quantity compared to $\omega_{qn}$.

If the qubit circuit possesses adequate anharmonicity, and all the control
pulses involved are carefully designed to avoid the excitation to higher
energy levels, then the Hamiltonian can be confined to the single-particle
case, thus arriving at
\begin{equation}
\hat{H}_{\text{eff}}=\hbar\omega_{q1}\hat{\sigma}_{11}+\hbar\eta_{1}%
\cos\left(  \tilde{\delta}t+\phi_{0}\right)  \hat{\sigma}_{11}.
\end{equation}
If we further focus on the flux qubit circuit which is typically treated as an
ideal two-level system where $\delta_{11}=\infty$, we have a simple result
$\eta_{1}\approx-\frac{\left\vert \Omega_{10}\Omega_{20}\right\vert }%
{\delta_{10}}$ and then $\hat{H}_{\text{eff}}$ becomes the form of Eq.%
%TCIMACRO{\TeXButton{TeX field}{~}}%
%BeginExpansion
~%
%EndExpansion
(\ref{ea:Hq_modulated}).

Now, we discuss the limit that the anharmonicity of the qubit is so weak that
Eq.%
%TCIMACRO{\TeXButton{TeX field}{~}}%
%BeginExpansion
~%
%EndExpansion
(\ref{eq:Hq_append}) becomes the form of a driven resonator. In this case, the
parameters can be represented as $\omega_{n}=n\bar{\omega}$, $\Omega
_{jn}=\sqrt{n+1}\bar{\Omega}_{j}$, and $\delta_{jn}=Const,$ where $\bar
{\omega}$ is the fundamental frequency of the resonator and $\bar{\Omega}_{j}$
is the driving strength on the resonator. Using such parameters, one can
obtain that the Stark shift $\nu_{n}=0$ and $\eta_{n}=0$, and thus the
periodical modulation of the qubit frequency vanishes. Therefore, to achieve
the periodical modulation using two-tone driving fields, the superconducting
qubit circuit should maintain a nonzero anharmonicity. In principle, the
periodical modulation effect shall exist only if the anharmonicity of the
interested qubit circuit is nonzero. This character requires a wider
anharmonicity range of the qubit circuit than in Ref.%
%TCIMACRO{\TeXButton{TeX field}{~}}%
%BeginExpansion
~%
%EndExpansion
\cite{Alaeian2019PRA}, where the anharmonicity of the transmon qubit circuit
needs to be negligibly small. Since the nonlinearity is a key factor for
demonstrating quantum phenomena%
%TCIMACRO{\TeXButton{TeX field}{~}}%
%BeginExpansion
~%
%EndExpansion
\cite{Wendin2007LTP}, we think periodically modulating the qubit circuit with
better anharmonicity is significant for exploring nonequilibrium quantum physics.

\section{Treatment into the interaction picture\label{append:Treatment}}

The full Hamiltonian with periodically modulated qubit frequency is given by
\begin{align}
\hat{H}_{\text{f}}=  &  \sum_{l}\sum_{d=\text{L,R}}\left[  \frac{\hbar}%
{2}\omega_{d}\sigma_{z}^{\left(  d,l\right)  }-\frac{\hbar}{2}\Omega
\cos\left(  \delta t+\phi_{d,l}\right)  \sigma_{z}^{\left(  \text{L},l\right)
}\right] \nonumber\\
&  -\sum_{l}\hbar g\sigma_{-}^{\left(  d,l\right)  }\sigma_{+}^{\left(
d,l+1\right)  }+\text{H.c.},\nonumber\\
&  -\sum_{l}\hbar K\sigma_{-}^{\left(  \text{L},l\right)  }\sigma_{+}^{\left(
\text{R},l\right)  }+\text{H.c.}%
\end{align}
where the subscript L and R represent the left and right legs of the ladder,
$l$ the lattice site, $\omega_{d}$ ($d=\mathrm{L,R}$) the qubit frequency on
the leg $d$, $g$ the intraleg tunneling rate, and $K$ the interleg tunneling
rate. To eliminate the time-dependent terms in Eq.%
%TCIMACRO{\TeXButton{TeX field}{~}}%
%BeginExpansion
~%
%EndExpansion
(\ref{eq:H}), we now apply to Eq.%
%TCIMACRO{\TeXButton{TeX field}{~}}%
%BeginExpansion
~%
%EndExpansion
(\ref{eq:H}) a unitary transformation $U_{d}\left(  t\right)  =%
%TCIMACRO{\tprod \limits_{l}}%
%BeginExpansion
{\textstyle\prod\limits_{l}}
%EndExpansion%
%TCIMACRO{\tprod \limits_{d=\mathrm{L,R}}}%
%BeginExpansion
{\textstyle\prod\limits_{d=\mathrm{L,R}}}
%EndExpansion
\exp\left[  iF_{l,d}\left(  t\right)  \right]  $ with%
\begin{equation}
F_{l,d}\left(  t\right)  =\frac{\sigma_{z}^{\left(  d,l\right)  }}{2}\left[
\frac{\Omega}{\delta}\sin\left(  \delta t+\phi_{d,l}\right)  +\omega
_{d}t\right]  ,
\end{equation}
in which manner, we now enter the interaction picture, and obtain the
effective Hamiltonian as
\begin{align}
\hat{H}_{\text{f}}=  &  -\sum_{l}\sum_{d=\mathrm{L,R}}\left[  \hbar
g\sigma_{-}^{\left(  \text{L},l\right)  }\sigma_{+}^{\left(  \text{L}%
,l+1\right)  }e^{i\alpha_{\text{L},l}\left(  t\right)  }+\text{H.c.}\right]
\nonumber\\
&  -\sum_{l}\sum_{d=\mathrm{L,R}}\left[  \hbar g\sigma_{-}^{\left(
\text{L},l\right)  }\sigma_{+}^{\left(  \text{L},l+1\right)  }e^{i\alpha
_{\text{R},l}\left(  t\right)  }+\text{H.c.}\right] \nonumber\\
&  -\sum_{l}\left[  \hbar K\sigma_{-}^{\left(  \text{L},l\right)  }\sigma
_{+}^{\left(  \text{R},l\right)  }e^{i\beta_{l}\left(  t\right)  }%
+\text{H.c.}\right]  \text{.}%
\end{align}
Here, the phase parameters $\alpha_{d,l}\left(  t\right)  $ and $\beta
_{l}\left(  t\right)  $ are
\begin{align}
\alpha_{d,l}\left(  t\right)   &  =\left[  \frac{2\Omega}{\delta}\sin
\phi_{d,l}^{\left(  -\right)  }\right]  \cos\left(  \delta t+\phi
_{d,l}^{\left(  +\right)  }\right)  ,d=\text{L,R}\\
\beta_{l}\left(  t\right)   &  =\left[  \frac{2\Omega}{\delta}\sin\phi
_{l}^{\left(  -\right)  }\right]  \cos\left(  \delta t+\phi_{l}^{\left(
+\right)  }\right)  +\Delta,
\end{align}
where $\phi_{d,l}^{\left(  \pm\right)  }=\left(  \phi_{d\text{,}l}\pm
\phi_{d\text{,}l+1}\right)  /2$, $\phi_{l}^{\left(  \pm\right)  }=\left(
\phi_{\text{L,}l}\pm\phi_{\text{R,}l+1}\right)  /2$, and $\Delta
=\omega_{\text{R}}-\omega_{\text{L}}$ is the qubit frequency difference
between different legs. Furthermore, we define $\phi_{d\text{,}l}=\phi
_{d}-\phi l$, $\phi_{\text{L}}=-\phi_{\text{R}}=\phi_{0}$, and use the
relation $\exp\left(  ix\sin\theta\right)  =\sum_{n}J_{n}\left(  x\right)
e^{in\theta}$, where $J_{n}\left(  x\right)  $ is the $n$th Bessel function of
the first kind, which yields the Hamiltonian as%
%TCIMACRO{\TeXButton{TeX field}{~}}%
%BeginExpansion
~%
%EndExpansion
\cite{Liu2014NJP,Zhao2015PRA}
\begin{align}
\hat{H}_{\text{f}}^{\prime}=  &  -\sum_{ln}\hbar g_{0}\sigma_{-}^{\left(
\text{L},l\right)  }\sigma_{+}^{\left(  \text{L},l+1\right)  }J_{xnl}^{\left(
+\right)  }\left(  t\right)  +\text{H.c.}\nonumber\\
&  -\sum_{ln}\hbar g_{0}\sigma_{-}^{\left(  \text{R},l\right)  }\sigma
_{+}^{\left(  \text{R},l+1\right)  }J_{xnl}^{\left(  -\right)  }\left(
t\right)  +\text{H.c.}\nonumber\\
&  -\sum_{ln}\hbar K_{0}\sigma_{-}^{\left(  \text{L},l\right)  }\sigma
_{+}^{\left(  \text{R},l\right)  }J_{ynl}\left(  t\right)  +\text{H.c..}%
\end{align}
Here, the parameters $J_{xnl}^{\left(  \pm\right)  }\left(  t\right)  $ and
$J_{ynl}\left(  t\right)  $ can be explicitly given by%
\begin{align}
J_{xnl}^{\left(  \pm\right)  }  &  =i^{N}J_{n}\left(  \eta_{x}\right)
\exp\left[  in\left(  \delta t\pm\phi_{0}-\phi l-\frac{\phi}{2}\right)
\right]  ,\\
J_{ynl}  &  =i^{N}J_{n}\left(  \eta_{y}\right)  \exp\left[  in\left(  \delta
t-\phi l\right)  +i\Delta t\right]  .
\end{align}
where $\eta_{x}=\frac{2\Omega}{\delta}\sin\left(  \frac{\phi}{2}\right)  $,
$\eta_{y}=\frac{2\Omega}{\delta}\sin\left(  \phi_{0}\right)  $, and
$J_{n}\left(  \cdot\right)  $ is the Bessel function of the first kind. We now
assume the detuning $\delta$ is tuned to match $\Delta$, i.e., $\delta=\Delta
$, such that, neglecting fast-oscillating terms, we can obtain the effective
Hamiltonian
\begin{align}
\hat{H}_{\text{ld}}=  &  -\sum_{l}\sum_{d=\text{L,R}}\hbar g\sigma
_{-}^{\left(  d,l\right)  }\sigma_{+}^{\left(  d,l+1\right)  }+\text{H.c.}%
\nonumber\\
&  -\sum_{l}\hbar K\sigma_{-}^{\left(  \text{L},l\right)  }\sigma_{+}^{\left(
\text{R},l\right)  }\exp\left(  i\phi l\right)  +\text{H.c.,}%
\end{align}
where $g=g_{0}J_{0}\left(  \eta_{x}\right)  $ and $K=K_{0}J_{1}\left(
\eta_{y}\right)  $ can be tunable in principle via modifying the two-tone
driving strength $\Omega$.

\section{Exact solution of the fidelity with the
environment\label{Append:exact}}

As the main text demonstrates, the effect of the environment on the state
generation process can be described by the Lindblad master equation
\begin{equation}
\frac{\text{d}\hat{\rho}}{\text{d}t}=\frac{1}{i\hbar}\left[  \hat
{H}_{\text{ld}}^{\left(  N\right)  }+\hat{H}_{\text{ld,g}}^{\prime},\hat{\rho
}\right]  +\mathcal{L}_{\mu1}\left[  \hat{\rho}\right]  . \label{eq:ME_append}%
\end{equation}
Here, $\hat{\rho}$ is the density operator of the ladder, $\mathcal{L}_{\mu
1}\left[  \hat{\rho}\right]  $ represents the Lindblad dissipation terms as%
\begin{align}
\mathcal{L}_{\mu1}\left[  \hat{\rho}\right]   &  =-\gamma_{1}\left\vert
\mu_{1}\right\rangle
%TCIMACRO{\TeXButton{TeX field}{\!}}%
%BeginExpansion
\!%
%EndExpansion
\left\langle \mu_{1}\right\vert \left\langle \mu_{1}\right\vert
%TCIMACRO{\TeXButton{TeX field}{\!}}%
%BeginExpansion
\!%
%EndExpansion
\hat{\rho}%
%TCIMACRO{\TeXButton{TeX field}{\!}}%
%BeginExpansion
\!%
%EndExpansion
\left\vert \mu_{1}\right\rangle
%TCIMACRO{\TeXButton{TeX field}{\!}}%
%BeginExpansion
\!%
%EndExpansion
+%
%TCIMACRO{\TeXButton{TeX field}{\!}}%
%BeginExpansion
\!%
%EndExpansion
\gamma_{1}\left\vert 0\right\rangle
%TCIMACRO{\TeXButton{TeX field}{\!}}%
%BeginExpansion
\!%
%EndExpansion
\left\langle 0\right\vert \left\langle 0\right\vert
%TCIMACRO{\TeXButton{TeX field}{\!}}%
%BeginExpansion
\!%
%EndExpansion
\hat{\rho}%
%TCIMACRO{\TeXButton{TeX field}{\!}}%
%BeginExpansion
\!%
%EndExpansion
\left\vert 0\right\rangle
%TCIMACRO{\TeXButton{TeX field}{\!}}%
%BeginExpansion
\!%
%EndExpansion
\nonumber\\
&  -%
%TCIMACRO{\TeXButton{TeX field}{\!}}%
%BeginExpansion
\!%
%EndExpansion
\frac{\Gamma_{1}}{2}\left\vert \mu_{1}\right\rangle
%TCIMACRO{\TeXButton{TeX field}{\!}}%
%BeginExpansion
\!%
%EndExpansion
\left\langle 0\right\vert \left\langle \mu_{1}\right\vert
%TCIMACRO{\TeXButton{TeX field}{\!}}%
%BeginExpansion
\!%
%EndExpansion
\hat{\rho}%
%TCIMACRO{\TeXButton{TeX field}{\!}}%
%BeginExpansion
\!%
%EndExpansion
\left\vert 0\right\rangle
%TCIMACRO{\TeXButton{TeX field}{\!}}%
%BeginExpansion
\!%
%EndExpansion
-%
%TCIMACRO{\TeXButton{TeX field}{\!}}%
%BeginExpansion
\!%
%EndExpansion
\frac{\Gamma_{1}}{2}\left\vert \mu_{1}\right\rangle
%TCIMACRO{\TeXButton{TeX field}{\!}}%
%BeginExpansion
\!%
%EndExpansion
\left\langle 0\right\vert \left\langle \mu_{1}\right\vert
%TCIMACRO{\TeXButton{TeX field}{\!}}%
%BeginExpansion
\!%
%EndExpansion
\hat{\rho}%
%TCIMACRO{\TeXButton{TeX field}{\!}}%
%BeginExpansion
\!%
%EndExpansion
\left\vert 0\right\rangle ,
\end{align}
and $\gamma_{1}$ ($\Gamma_{1}$) is the relaxation (dephasing) rate of the
single-particle ground state $\left\vert \mu_{1}\right\rangle $. Solving Eq.%
%TCIMACRO{\TeXButton{TeX field}{~}}%
%BeginExpansion
~%
%EndExpansion
(\ref{eq:ME_append}), where the Hilbert space is $\left\{  \left\vert
0\right\rangle ,\left\vert \mu_{1}\right\rangle \right\}  $, we can obtain the
population on $\left\vert \mu_{1}\right\rangle $ after some time $t$, i.e.,
\begin{align}
\rho_{11}  &  =\left\langle \mu_{1}\right\vert
%TCIMACRO{\TeXButton{TeX field}{\!}}%
%BeginExpansion
\!%
%EndExpansion
\hat{\rho}%
%TCIMACRO{\TeXButton{TeX field}{\!}}%
%BeginExpansion
\!%
%EndExpansion
\left\vert \mu_{1}\right\rangle \nonumber\\
&  =r_{0}-r_{0}\operatorname{Re}\left\{  \left(  1-\frac{i\gamma_{1}^{\prime}%
}{2C_{1}^{\prime}}\right)  e^{-\frac{1}{2}\gamma_{1}^{\prime}t}\exp\left(
itC_{1}^{\prime}\right)  \right\}
%TCIMACRO{\TeXButton{TeX field}{\!}}%
%BeginExpansion
\!%
%EndExpansion
.
\end{align}
Here, the intermediate parameters are explicitly given as follows,
\begin{align}
r_{0}  &  =\frac{\frac{C_{1}^{2}}{2}}{C_{1}^{2}+\frac{\gamma_{1}\Gamma_{1}}%
{2}},\\
C_{1}^{\prime}  &  =\sqrt{C_{1}^{2}-\frac{1}{4}\left(  \gamma_{1}-\frac
{\Gamma_{1}}{2}\right)  ^{2}},\\
\gamma_{1}^{\prime}  &  =\gamma_{1}+\frac{\Gamma_{1}}{2},
\end{align}
and $\rho_{11}$ is also called the fidelity of $\left\vert \mu_{1}%
\right\rangle $. In the limit of strong coupling ($C_{1}\gg\gamma_{1}$,
$\Gamma_{1}$), $r_{0}=\frac{1}{2}$, $C_{1}^{\prime}=C_{1}$, and $\gamma
_{1}^{\prime}/C_{1}^{\prime}=0$, thus yielding%
\begin{equation}
\rho_{11}=\frac{1}{2}\left[  1-e^{-\frac{1}{2}\gamma_{1}^{\prime}t}\cos\left(
C_{1}t\right)  \right]  ,
\end{equation}
which yields $\rho_{11}=\frac{1}{2}$ in the steady state ($t=\infty$).

\end{document}